\documentclass[10pt,twoside]{article}
\usepackage[T1]{fontenc}
\usepackage[utf8]{inputenc}
\usepackage[british]{babel}
\usepackage{amsmath,amsthm,mathtools,bm}
\IfFileExists{newtxtext.sty}{\usepackage{newtxtext,newtxmath}}{\usepackage{txfonts}}
\usepackage[paperwidth=6.75in,paperheight=9.75in,
  left=0.72in,right=0.72in,top=0.78in,bottom=0.72in,
  headheight=13pt,headsep=16pt,footskip=23pt]{geometry}
\usepackage{booktabs,tabularx,array}
\usepackage{tikz,pgfplots}
\pgfplotsset{compat=1.18}
\usepgfplotslibrary{fillbetween}
\usepackage{natbib}
\usepackage{microtype}
\usepackage{needspace,float}
\usepackage{setspace}
\usepackage{enumitem}
\usepackage{titlesec}
\usepackage{fancyhdr}
\usepackage{hyperref}
\hypersetup{colorlinks=true,linkcolor=black,citecolor=black,urlcolor=black,
 pdftitle={Strategic Repayment and Latent Fragility in Financial Networks},
 pdfauthor={Hicham El Kaissi},pdfsubject={University of Kent theoretical working paper}}
\renewcommand{\thesection}{\Roman{section}}

\renewcommand{\theequation}{\arabic{equation}}
\newtheorem{proposition}{Proposition}
\newtheorem{lemma}[proposition]{Lemma}
\newtheorem{corollary}[proposition]{Corollary}
\newtheorem{assumption}[proposition]{Assumption}
\theoremstyle{remark}
\newtheorem{remark}[proposition]{Remark}
\newtheorem{definition}[proposition]{Definition}
\titleformat{\section}[block]{\normalfont\normalsize\centering}{\thesection.}{0.55em}{}
\titlespacing*{\section}{0pt}{16pt plus 3pt minus 2pt}{9pt plus 2pt minus 1pt}
\titleformat{\subsection}[block]{\normalfont\itshape\centering}{\Alph{subsection}.}{0.55em}{}
\titlespacing*{\subsection}{0pt}{13pt plus 3pt minus 2pt}{7pt plus 1pt minus 1pt}
\titleformat{\paragraph}[runin]{\normalfont\itshape}{ }{0pt}{}[.]
\titlespacing*{\paragraph}{0pt}{9pt plus 2pt minus 1pt}{0.5em}
\fancypagestyle{plain}{\fancyhf{}\fancyfoot[C]{\scriptsize\thepage}}
\allowdisplaybreaks[2]
\newcommand{\EE}{\mathbb E}
\newcommand{\PP}{\mathbb P}
\newcommand{\RR}{\mathbb R}
\newcommand{\one}{\mathbf 1}
\newcommand{\pos}[1]{\left[#1\right]^+}
\newcommand{\clip}{\operatorname{clip}_{[0,1]}}
\newcommand{\pb}{\bar p}
\newcommand{\pbtot}{\bar p_{\mathrm{tot}}}
\newcommand{\zb}{\bar z}
\newcommand{\A}{\mathcal A}
\newcommand{\ind}{\mathbf 1}
\newcommand{\EN}{\mathrm{EN}}
\newcommand{\G}{\mathrm G}

\begin{document}
\thispagestyle{plain}
{\noindent\footnotesize\itshape Working Paper\hfill November 2026\par}
\vspace*{0.50in}
\begin{center}
{\fontsize{14.3}{17.5}\selectfont Strategic Repayment and Latent Fragility in Financial Networks:\\[1pt]Why Banks Delay Payments They Can Make\par}
\vspace{0.24in}
{\normalsize By \textsc{Hicham El Kaissi}\footnotemark\par}
\vspace{0.24in}
\end{center}
\footnotetext{Hicham El Kaissi: University of Kent
(email: \href{mailto:h.el-kaissi@kent.ac.uk}{h.el-kaissi@kent.ac.uk}).}
\begin{center}\begin{minipage}{0.88\textwidth}\itshape\noindent
Banks can withhold payable cash to cover funding needs before deferred principal falls due. Under scale-family shocks with an affine scale response, optimal reserve choice yields a piecewise-affine clearing rule with two reserve parameters per bank. When missed receipts raise desired reserves, a bank whose repayment remains interior cuts outgoing payments by more than it misses. This feedback can amplify delays even in a unique equilibrium. We derive exact conditions for full early settlement and a complete freeze to coexist when Eisenberg--Noe clearing uniquely predicts full payment. Full settlement alone cannot reveal this fragility because the reserve response is inactive when every receipt arrives. Coexistence depends only on bank-level payment totals, opening cash and reserve parameters. Rewiring at fixed totals can change propagation and partial freezes. In a four-bank illustration, a partial freeze delays \pounds233 million of \pounds255 million due, and every obligation ultimately settles.

\end{minipage}\end{center}
\begin{center}\begin{minipage}{0.88\textwidth}
{\small\noindent\textit{(JEL C72, D85, E42, G01, G21)}\par}
\vspace{3pt}
{\footnotesize\noindent\textit{Keywords:} financial networks; precautionary liquidity;
strategic repayment; payment clearing; coordination failure; systemic risk.\par}
\end{minipage}\end{center}

\section{Introduction}\label{sec:intro}
A bank can miss a payment even when it has enough cash to make it. When banks expect less from their counterparties, retaining cash for a later funding need can be individually optimal and make low settlement self-fulfilling. We study when a bank delays a payment it can afford and how that decision propagates through its creditors. Under the proportional-clearing benchmark of \citet{EisenbergNoe2001}, each bank remits as much as contemporaneous resources permit.

Retained cash insures a funding need that arises after early settlement and before deferred obligations fall due. Deferral incurs contractual late interest and a private continuation-value cost. Section~\ref{subsec:arrears} shows that positive deferral costs restore maximal feasible early payment when the funding need and deferred principal fall due together and retained cash has no earlier use.

Stochastic optimisation gives a reserve target equal to a critical quantile of the funding need. At fixed obligations and opening cash, the baseline reserve $h_i$ and its sensitivity to missed receipts $\nu_i$ determine the clearing rule for every shock law satisfying Assumption~\ref{ass:scale}. The affine funding-scale response yields piecewise-affine best responses, an exact mixed complementarity formulation and a Tarski lattice of equilibria. Appendix~\ref{app:distribution} extends the representation to peaks-over-threshold shocks with a fixed exceedance probability.

While a bank's repayment remains interior, a pound of missed receipts drains a pound of cash and raises desired reserves by $\nu_i$ pounds. Its outgoing payments fall by $1+\nu_i$ pounds. The dealer--client example shows how this feedback can amplify delays in a unique equilibrium at fixed asset prices, with all debts ultimately paid.

We derive exact conditions for a system-wide freeze to coexist with full payment. The freeze margin $\Gamma_0$ flags whether an all-zero equilibrium is possible even while a network settles in full. It uses bank-level claim and obligation totals, opening cash and reserve parameters. With those inputs held fixed, rewiring can change propagation and sustain additional partial freezes. In the four-bank example, concentrating the links at fixed bank-level totals raises the number of equilibria from three to nine. With common payment totals and behavioural parameters, the coexistence frontier depends on the reserve quantile relative to mean funding need and takes closed form for generalised Pareto shocks.

The repayment rule separates deliberate withholding from capacity lost when counterparties pay less. The bank that yields the largest marginal gain in settlement can differ from the bank that ends a freeze with the smallest injection. An upstream bank can be the cheaper target when its additional payment supplies cash and lowers the receiving dealer's reserve demand. The margin-window example compares funded late charges, early access to arrears, cash injections and claim prefunding on the same network.

The mechanism links payment-timing incentives \citep{BechGarratt2003,Bech2008} to private reserve choice \citep{Edgeworth1888,AcharyaSkeie2011,HeiderEtAl2015,Su2018,DenbeeEtAl2021}. In a Fedwire-calibrated model, \citet{AfonsoShin2011} show how cautious payment rules can reinforce one another. We derive the repayment rule from optimal reserve choice and use it to characterise exactly when the same liabilities and opening cash support both full early settlement and a freeze. Related network work studies self-fulfilling insolvencies, dependency cycles and bailout policies \citep{JacksonPernoud2024} and strategic payments and default \citep{BertschingerHoeferSchmand2020,AllouchJalloulDuncan2023}. The payment multiplier connects to spectral interaction and crisis amplification \citep{BallesterEtAl2006,BramoulleEtAl2014,GaiHaldaneKapadia2011,BrunnermeierReis2023}.

Evidence on precautionary reserve holding during the 2007--08 turmoil \citep{AshcraftMcAndrewsSkeie2011} and later Fedwire payment timing \citep{ArmantierArnoldMcAndrews2008} motivates the liquidity and timing frictions.

\section{The Model}\label{sec:model}

\subsection{Timing, payment links and feasible repayment}
Time is normalised to two settlement dates. At $t=0$, banks simultaneously choose early payments with complete information about the network and economic primitives. A temporary funding need arises during $0<t<1$, after early settlement but before deferred obligations fall due at $t=1$. Retained cash provides liquidity services during this interval and is released before final settlement. Section~\ref{subsec:terminal} specifies the terminal accounts. Best responses at $t=0$ condition on conjectured counterparty payments; they do not require sequential observation within the settlement window.

Let $\mathcal N=\{1,\ldots,N\}$ be the institutions and $L\in\RR_+^{N\times N}$ the fixed nominal liability matrix, with $L_{ii}=0$. A directed link $i\to j$ carries the amount $L_{ij}$ owed by debtor $i$ to creditor $j$, including interest accrued before date 0. Define total outgoing obligations, the active debtor set and relative liabilities by
\begin{equation}\label{eq:network}
\pb_i=\pb_i(L):=\sum_j L_{ij},\qquad \A=\A(L):=\{i:\pb_i>0\},\qquad
\Pi_{ij}=\Pi_{ij}(L):=\frac{L_{ij}}{\pb_i}\quad(i\in\A).
\end{equation}
We assume $\A\ne\varnothing$ and set the rows of $\Pi$ to zero outside $\A$. Active bank $i$ chooses a repayment fraction $s_i\in[0,1]$. Its total and bilateral early payments are
\begin{equation}\label{eq:payments}
p_i^0(s_i)=\pb_i s_i,\qquad
p_{ij}^0(s_i)=L_{ij}s_i=\Pi_{ij}p_i^0(s_i).
\end{equation}
We abbreviate $p_i^0(s_i)$ by $p_i(s_i)$ for date-0 flows. Pure creditors outside $\A$ have no repayment choice. In expressions over all institutions, extend the active profile $s=(s_i)_{i\in\A}$ by setting $s_i=p_i=0$ outside $\A$. Incoming cash, contractual claims and missing inflows are
\begin{equation}\label{eq:inflows}
\begin{aligned}
z_i(s_{-i})&=\sum_{j\ne i}L_{ji}s_j
=\bigl(\Pi^\top p(s)\bigr)_i,\\
\zb_i&=\sum_{j\ne i}L_{ji},\qquad
\Delta_i(s_{-i})=\zb_i-z_i(s_{-i})\ge0.
\end{aligned}
\end{equation}
Thus $z_i$ depends only on debtors linked to $i$, and is independent of $s_i$. For $\zb_i>0$, define $m_i(s_{-i})=z_i(s_{-i})/\zb_i$; set $m_i=0$ if $\zb_i=0$.

Let $e_i\in\RR$ be net unrestricted date-0 liquid resources outside the modelled network, after senior near-term outside obligations. This excludes cash pledged for late interest in Section~\ref{subsec:discipline}. Available cash and retained reserves are
\begin{equation}\label{eq:resources}
a_i(s_{-i})=\pos{e_i+z_i(s_{-i})},\qquad
R_i(s_i,s_{-i})=a_i(s_{-i})-p_i(s_i)\ge0.
\end{equation}
If $e_i+z_i<0$, no cash is available for interbank settlement; any remaining outside shortfall is outside the settlement account. The feasible choice set is
\begin{equation}\label{eq:feasible}
\mathcal S_i(s_{-i})=\left[0,\bar s_i(s_{-i})\right],\qquad
\bar s_i(s_{-i}):=\min\left\{1,\frac{a_i(s_{-i})}{\pb_i}\right\}.
\end{equation}
A \emph{generalised Nash game} allows each player's feasible actions, as well as payoffs, to depend on the other players' choices. That formulation is needed here because incoming payments finance outgoing settlement. For example, if $\pb_i=100$, $e_i=20$ and one debtor owes $L_{ji}=80$, then $\bar s_i(s_j)=0.2+0.8s_j$: full repayment is feasible when $s_j=1$, whereas only 20\% is feasible when $s_j=0$. An equilibrium requires each bank to choose its best feasible action at the same counterparty profile.

\begin{definition}[Maximal feasible or mechanical repayment]\label{def:mechanical}
Mechanical repayment is $\Phi_i(s_{-i})=\bar s_i(s_{-i})$. A mechanical clearing vector satisfies $s=\Phi(s)$. With $e_i\ge0$ and proportional-payment priority, this is the Eisenberg--Noe operator in repayment fractions. With negative $e_i$, it is an adapted \emph{EN-type} operator.
\end{definition}
The strategic model characterises when a bank retains some of the cash it could remit under this benchmark.

\subsection{Future liquidity risk and confidence-dependent stress}
The interim funding need may arise from margin calls, a temporary failure to roll borrowing or a bridge before scheduled receipts arrive. A missed incoming payment can tighten funding conditions as well as reduce current cash. We capture the later effect through the scale of a distinct temporary requirement; the missing contractual inflow is already included in $a_i$.

For example, suppose $i$ is owed 80, has baseline funding-need scale 20 and scale pass-through 0.5. Full counterparty payment leaves the scale at 20. If its debtor pays 60, current cash falls by 20 and the later-need scale rises to $20+0.5(80-60)=30$. With a unit-mean exponential normalised shock, the corresponding expected interim need rises from 20 to 30. The following assumption generalises this example while keeping the normalised shock distribution fixed.

\Needspace{10\baselineskip}
\begin{assumption}[Scale-family liquidity need]\label{ass:scale}
Conditional on counterparty repayments, active bank $i$ faces
\begin{equation}\label{eq:scale}
\Lambda_i(s_{-i},Y_i)=\kappa_i(s_{-i})Y_i,\qquad
\kappa_i(s_{-i})=\omega_i+\psi_i\Delta_i(s_{-i}),\quad
\omega_i>0,\quad\psi_i\ge0.
\end{equation}
The nonnegative normalised shock $Y_i$ has a continuous strictly decreasing survival function $\bar G_i(x)=\PP(Y_i>x)$, with $\bar G_i(0)=1$, $\lim_{x\to\infty}\bar G_i(x)=0$, a continuous strictly positive density $f_i(x)=-\bar G_i'(x)$ on $x\ge0$, and $\EE[Y_i]<\infty$. Its law is invariant to $s_{-i}$. The dimensionless coefficient $\psi_i$ is fixed across payment profiles.
\end{assumption}
The scale $\kappa_i$ measures future funding need. Each missing unit raises it by $\psi_i$; only incoming links affect this response. With $m_i^Y:=\EE[Y_i]$, expected need is $\kappa_i(s_{-i})m_i^Y$, and its pass-through per unit of missing inflow is
\begin{equation}\label{eq:meanpass}
\eta_i:=\frac{\partial\EE[\Lambda_i\mid s_{-i}]}{\partial\Delta_i}
=\psi_i m_i^Y.
\end{equation}
The model allows any $\eta_i\ge0$. Section~\ref{sec:illustrations} derives the pass-through threshold for coexistence over a positive-width range of external liquidity buffers and tests whether alternative bounds permit pass-through above that threshold. A unit bound allows up to one extra unit of expected interim funding need per unit of missing settlement. The illustrations use a full-distribution generalised Pareto law. Appendix~\ref{app:distribution} gives a peaks-over-threshold extension with a possible atom at zero.

\subsection{Economic discipline and precautionary liquidity valuation}\label{subsec:discipline}
Each unit of interim need uncovered by reserves entails an incremental funding cost $\chi_i>0$. When a bank can bridge the shortage by borrowing, a simple-interest calibration over the funding horizon gives
\[
\chi_i \simeq (r_i^{L}-r_i^{R})\tau_i^{\Lambda}+\chi_i^{o},
\]
where $r_i^{L}$ is the annualised marginal bridge-borrowing rate, $r_i^{R}$ is the annualised return forgone when retained cash is used instead, $\tau_i^{\Lambda}$ is the bridge duration in years, and $\chi_i^{o}$ captures other incremental costs per unit. Distressed asset sales provide another source of funding costs \citep{ShleiferVishny1992,BrunnermeierPedersen2009}.

Deferring one unit costs the bank both contractual late interest and a private continuation-value loss. With annualised simple late rate $r_i^{\mathrm{late}}\ge0$ and delay window $\tau_i^d>0$ in years, the total marginal cost is
\begin{equation}\label{eq:delaycost}
\delta_i=d_i^{\mathrm{priv}}+k_i>0,\qquad
k_i:=r_i^{\mathrm{late}}\tau_i^d .
\end{equation}
The rate is common across bank $i$'s creditors. The term $d_i^{\mathrm{priv}}$ captures reputation, franchise value and market access net of any carry benefit, with each carry cost counted once. Both components are fixed across repayment profiles.

The contract backs late interest with debtor-owned cash $C_i\ge k_i\pb_i$ pledged before date 0. The pledge covers the charge even if the bank defers all its obligations. It is reserved for interest until after final settlement and excluded from available settlement cash at both dates. For deferred principal $A_i^1(s_i):=\pb_i(1-s_i)$, the bank pays interest $I_i^1(s_i):=k_i A_i^1(s_i)$. Its combined deferral cost is $d_i^{\mathrm{priv}}A_i^1(s_i)+I_i^1(s_i)=\delta_i A_i^1(s_i)$.

Writing $s=(s_i,s_{-i})$, the bank maximises
\begin{equation}\label{eq:netU}
\begin{aligned}
U_i^{\mathrm{net}}(s_i,s_{-i})
={}&-\chi_i\EE\!\left[(\kappa_i(s_{-i})Y_i-R_i(s))^+\mid s_{-i}\right]\\
&-\delta_i\pb_i(1-s_i),\qquad s_i\in\mathcal S_i(s_{-i}).
\end{aligned}
\end{equation}
The finite-mean assumption makes this payoff finite. An equivalent objective for the bank's own choice is
\begin{equation}\label{eq:grossU}
\begin{aligned}
U_i(s_i,s_{-i})&=V_i(R_i(s);\kappa_i(s_{-i}))-\delta_i\pb_i(1-s_i),\\
V_i(R;\kappa)&=\chi_i\EE[\min\{\kappa Y_i,R\}]
=\chi_i\int_0^R\bar G_i(x/\kappa)\,dx.
\end{aligned}
\end{equation}
The difference $U_i-U_i^{\mathrm{net}}=\chi_i\kappa_i(s_{-i})\EE[Y_i]$ is constant in $s_i$. Thus $U_i$ gives the same best response, while $U_i^{\mathrm{net}}$ permits comparisons across counterparty profiles. The liquidity-value derivatives are
\begin{equation}\label{eq:Vderiv}
V_{i,R}=\chi_i\bar G_i(R/\kappa),\qquad
V_{i,RR}=-\frac{\chi_i}{\kappa}f_i(R/\kappa)<0,\qquad
V_{i,R\kappa}=\frac{\chi_iR}{\kappa^2}f_i(R/\kappa)\ge0.
\end{equation}
All profile-dependent quantities are defined with their arguments; subsequent algebra suppresses arguments when the profile is fixed. The primitives $(L,e,\omega,\psi,\chi,\delta,\bar G)$ are held fixed unless a comparative-static change is specified.

\paragraph{Franchise costs of deferral}
If an unpaid fraction $1-s_i$ exposes franchise value $F_i>0$ to loss with severity $\lambda\vartheta>0$, the private component can be parameterised as
\begin{equation}\label{eq:franchise}
\mathrm{Disc}^{\mathrm F}_i(s_i)=\lambda\vartheta F_i(1-s_i),\qquad
d_i^{\mathrm{priv}}=\frac{\lambda\vartheta F_i}{\pb_i}
=\frac{\lambda\vartheta}{\upsilon_i},\quad
\upsilon_i:=\frac{\pb_i}{F_i}.
\end{equation}
This outgoing-obligations-to-franchise ratio is distinct from incoming exposure $\zb_i$. In the baseline, $\delta_i$ is fixed as $L$ varies.

\subsection{Final settlement or default}\label{subsec:terminal}
Deferred contractual claims remain enforceable. The settlement accounts follow the timing introduced at the start:
\begin{enumerate}[label=(\roman*),leftmargin=2.1em,itemsep=2pt]
\item \emph{Date 0:} bank $i$ settles $p_i^0(s_i)=\pb_i s_i$, retains $R_i(s)=a_i(s_{-i})-p_i^0(s_i)$, and leaves bilateral arrears $L_{ij}^1(s_i)=(1-s_i)L_{ij}$.
\item \emph{Interim, $0<t<1$:} reserves insure $\Lambda_i(s_{-i},Y_i)$. Cash used for this temporary bridge or posted as collateral returns before date 1. Funding premiums affect franchise payoff outside the terminal principal account.
\item \emph{Date 1:} restored reserves and unrestricted continuation cash $X_i\ge0$ fund deferred principal. The separate pledged account pays late interest. Neither principal nor interest receipts are recycled into the same settlement round.
\end{enumerate}
The conditional law $K_i^X(\cdot\mid s_{-i})$ of $X_i$ is independent of $s_i$. The variables $X_i$ and $Y_i$ may be correlated. Principal below denotes the fixed claim at date 0, including any interest accrued by then.

Late principal paid and principal in default are
\begin{equation}\label{eq:terminalrule}
\begin{aligned}
Q_i^1(s,X_i)&:=\min\{A_i^1(s_i),R_i(s)+X_i\},\\
\mathcal D_i^1(s,X_i)&:=A_i^1(s_i)-Q_i^1(s,X_i).
\end{aligned}
\end{equation}
The pledged account pays $I_i^1(s_i)$ in full. Creditor $j$ receives $\Pi_{ij}Q_i^1(s,X_i)$ in late principal and $\Pi_{ij}I_i^1(s_i)=k_i(1-s_i)L_{ij}$ in interest, and bears the principal loss $\Pi_{ij}\mathcal D_i^1(s,X_i)$. The unused pledge $C_i-I_i^1(s_i)$ returns to the debtor after settlement.

\begin{proposition}[Terminal settlement and timing neutrality of principal]\label{prop:terminal}
For every feasible date-0 profile and every realisation of $X_i\ge0$,
\begin{equation}\label{eq:terminalidentity}
\begin{aligned}
p_i^0(s_i)+Q_i^1(s,X_i)&=\min\{\pb_i,a_i(s_{-i})+X_i\},\\
\mathcal D_i^1(s,X_i)&=\pos{\pb_i-a_i(s_{-i})-X_i},\\
R_i(s)+X_i-Q_i^1(s,X_i)&=\pos{a_i(s_{-i})+X_i-\pb_i}.
\end{aligned}
\end{equation}
Conditional on $s_{-i}$, the bank's own early payment leaves total principal paid and principal default invariant. Total cash paid to creditors is $\min\{\pb_i,a_i(s_{-i})+X_i\}+I_i^1(s_i)$ and thus depends on early repayment when $k_i>0$. At a fixed realisation of $X_i$, counterparty payments affect principal default through $a_i(s_{-i})$.
\end{proposition}

For a nondecreasing default penalty $\mathrm{Pen}_i^{\mathrm{def}}:[0,\infty)\to[0,\infty)$ with $\mathrm{Pen}_i^{\mathrm{def}}(0)=0$, the completed debtor-side criterion is
\begin{equation}\label{eq:completepayoff}
\mathcal J_i(s_i,s_{-i})=U_i^{\mathrm{net}}(s_i,s_{-i})
-\EE\!\left[\mathrm{Pen}_i^{\mathrm{def}}(\mathcal D_i^1(s,X_i))\mid s_{-i}\right].
\end{equation}
The default penalty is bounded by $\mathrm{Pen}_i^{\mathrm{def}}(\pb_i)$. Proposition~\ref{prop:terminal} makes its expected value constant in $s_i$, so the completed criterion and $U_i^{\mathrm{net}}$ have the same unique best response. The criterion measures the bank's funding, deferral and own default costs. Creditor-side principal recoveries are outside this criterion. Incoming late interest $\sum_{j\in\A} k_j(1-s_j)L_{ji}$ is fixed in bank $i$'s own optimisation and can be added without changing its best response.

\section{Strategic Repayment and the Deviation from Clearing}\label{sec:individual}
\subsection{The critical-fractile retention rule}
Fix a bank $i\in\A$ and counterparties' actions $s_{-i}$. The bank's feasible reserve $R_i=a_i-\pb_i s_i$ falls by $\pb_i$ per unit increase in its own repayment fraction. From \eqref{eq:Vderiv},
\begin{equation}\label{eq:ownmarginal}
\frac{\partial U_i}{\partial s_i}=\pb_i\left[\delta_i-\chi_i \bar G_i(R_i/\kappa_i)\right],
\qquad \frac{\partial^2 U_i}{\partial s_i^2}
=-\frac{\chi_i\pb_i^2}{\kappa_i}f_i(R_i/\kappa_i)<0.
\end{equation}
The strictly concave problem has a unique solution. At an interior optimum, the marginal value of liquidity equals the marginal cost of deferral:
\begin{equation}\label{eq:fractile}
\chi_i\PP(\Lambda_i>R_i\mid s_{-i})=\delta_i.
\end{equation}
The target is a quantile of the funding-need distribution. Define the dimensionless critical retention ratio
\begin{equation}\label{eq:ci}
\begin{gathered}
c_i=c_i(\delta_i/\chi_i;\bar G_i):=\begin{cases}
\bar G_i^{-1}(\delta_i/\chi_i),&0<\delta_i<\chi_i,\\
0,&\delta_i\ge\chi_i
\end{cases},\\
h_i=h_i(\omega_i,c_i):=\omega_i c_i,
\qquad \nu_i=\nu_i(\psi_i,c_i):=\psi_i c_i.
\end{gathered}
\end{equation}
The ratio $c_i$ is strictly positive precisely when the funding shortage cost exceeds the marginal discipline cost. The (possibly unattainable) desired liquidity reserve is
\begin{equation}\label{eq:H}
H_i(s_{-i})=c_i\kappa_i(s_{-i})=h_i+\nu_i\Delta_i(s_{-i}).
\end{equation}
Thus the benchmark retention $h_i$ and confidence-response parameter $\nu_i$ are reduced-form sufficient statistics of the bank's behavioural response.

\begin{proposition}[Optimal strategic repayment]\label{prop:bestresponse}
Under Assumption~\ref{ass:scale}, the best response is unique and, for every counterparty profile, equals
\begin{equation}\label{eq:BR}
\begin{aligned}
B_i(s_{-i})
&=\clip\!\left(\frac{e_i+z_i(s_{-i})-H_i(s_{-i})}{\pb_i}\right)\\
&=\clip\!\left(\frac{e_i-h_i-\nu_i\zb_i+(1+\nu_i)z_i(s_{-i})}{\pb_i}\right).
\end{aligned}
\end{equation}
Here $\clip(x)=\min\{1,\max\{0,x\}\}$. The rule holds even when $e_i+z_i<0$. If $\delta_i\ge\chi_i$, then $c_i=h_i=\nu_i=0$ and $B_i=\Phi_i$ for all $s_{-i}$.
\end{proposition}

\subsection{The strategic repayment wedge}
The strategic repayment wedge is the difference between feasible and chosen payment. Define the repayment wedge and monetary withholding by
\begin{equation}\label{eq:wedge}
d_i(s_{-i})=\Phi_i(s_{-i})-B_i(s_{-i})\ge0,
\qquad w_i(s_{-i})=\pb_i d_i(s_{-i}).
\end{equation}
A bank with insufficient resources for full repayment may nevertheless have $w_i>0$: it withholds part of the amount it \emph{could} currently pay. The wedge measures delayed date-0 settlement. Proposition~\ref{prop:terminal} gives the associated terminal principal loss $\mathcal D_i^1$.

\begin{proposition}[Exact condition for a deviation from maximal feasible repayment]\label{prop:deviation}
Let $a_i(s_{-i})>0$ and define the liquidity remaining under the mechanical repayment decision by
\begin{equation}\label{eq:Rphi}
R_i^{\Phi}(s_{-i})=a_i(s_{-i})-\pb_i\Phi_i(s_{-i})
=\pos{a_i(s_{-i})-\pb_i}.
\end{equation}
Then $d_i(s_{-i})>0$ if and only if
\begin{equation}\label{eq:deviationtest}
\chi_i \bar G_i\!\left(R_i^{\Phi}/\kappa_i\right)>\delta_i.
\end{equation}
If $a_i=0$, neither mechanical nor strategic repayment is positive and $d_i=0$. In particular, when $\delta_i\ge\chi_i$ no feasible strategic deviation occurs, irrespective of the payment environment.
\end{proposition}
If $0<a_i<\pb_i$, maximal feasible payment leaves no reserves. The bank then withholds payable cash precisely when $\delta_i<\chi_i$. If $a_i\ge\pb_i$, it pays in full exactly when $a_i-\pb_i\ge H_i$.

\paragraph{Strategic withholding despite full payment capacity}
Consider a bank with $\pb_i=100$, $\zb_i=80$, $e_i=45$, baseline retention $h_i=7.6356$ and confidence coefficient $\nu_i=0.286335$. When all counterparties settle in full, $a_i=125$ and $B_i=1$. If incoming payments fall to $z_i=65$, the bank still has $a_i=110>100$ and thus \emph{can} settle all obligations. Its desired reserve, however, rises to $H_i=7.6356+0.286335(15)\simeq11.931$, yielding $p_i^B(s_{-i}):=\pb_iB_i(s_{-i})=a_i-H_i\simeq98.069$ and strictly positive strategic withholding of about 1.931 units.

Under the franchise-cost specification \eqref{eq:franchise}, $k_i\ge\chi_i$ suffices for maximal feasible repayment. If $k_i<\chi_i$, the condition $\delta_i\ge\chi_i$ becomes
\begin{equation}\label{eq:franchise_threshold}
\pb_i\le \frac{\lambda\vartheta F_i}{\chi_i-k_i}.
\end{equation}
At fixed franchise value and severity, the benefit from retaining a small interbank obligation can be outweighed by discipline. Changes in $L$ must also account for their effects on resources, exposures and the payment network.

\subsection{Two distinct counterparty channels}
When $e_i+z_i>0$ and holding $s_i$ fixed, the cash-flow identity gives
\begin{equation}\label{eq:account_derivative}
\frac{\partial R_i}{\partial s_j}=L_{ji},\qquad j\ne i.
\end{equation}

Optimal desired retention instead satisfies, from \eqref{eq:H},
\begin{equation}\label{eq:retention_derivative}
\frac{\partial H_i}{\partial s_j}=-\nu_iL_{ji},
\qquad\frac{\partial H_i}{\partial m_i}=-\nu_i\zb_i\quad(\zb_i>0).
\end{equation}
Define optimally retained cash by $R_i^B(s_{-i}):=a_i(s_{-i})-\pb_iB_i(s_{-i})$. In an interior repayment regime, $R_i^B=H_i$ and $\partial R_i^B/\partial s_j=-\nu_iL_{ji}$, while the accounting derivative \eqref{eq:account_derivative} is positive on an incoming link. If $\nu_i>0$, the bank passes along more than the extra cash received because the receipt also releases precautionary reserves.

\begin{proposition}[Strategic complementarity and confidence amplification]\label{prop:amplification}
Where $0<B_i<1$ so that clipping is inactive,
\begin{equation}\label{eq:amplification}
\frac{\partial B_i}{\partial s_j}=\frac{(1+\nu_i)L_{ji}}{\pb_i},\qquad
\frac{\partial(\pb_iB_i)}{\partial z_i}=1+\nu_i,\qquad
\frac{\partial B_i}{\partial m_i}=(1+\nu_i)\frac{\zb_i}{\pb_i}\quad(\zb_i>0).
\end{equation}
Moreover, for $j\ne i$ where $R_i(s)>0$, with $x_i(s):=R_i(s)/\kappa_i(s_{-i})$,
\begin{equation}\label{eq:crosspartial}
\frac{\partial^2 U_i}{\partial s_i\partial s_j}
=\frac{\chi_i\pb_iL_{ji}}{\kappa_i}f_i(x_i)
\left(1+\psi_ix_i\right)\ge0.
\end{equation}
The cross-partial generally varies with other banks' repayment choices, even though the interior best-response slope is constant under the affine scale specification. Where the unclipped response is strictly outside $[0,1]$, $B_i$ is locally flat; clipping thresholds require one-sided derivatives.
\end{proposition}

The game has strategic complementarities in repayment \citep{Topkis1979,MilgromRoberts1990}. On a direct link $j\to i$, an extra unit received raises interior outgoing payment by $1+\nu_i$: one unit from cash and $\nu_i$ from released precautionary reserves. If $L_{ji}=0$, changes in $s_j$ have no direct cash-flow or best-response effect on $i$; indirect effects travel through payment paths. For an unpaid share $1-m_i$, the increase in desired reserves is $\nu_i\zb_i(1-m_i)$. Appendix~\ref{app:nonlinear} extends the confidence channel to nonlinear stress.

\begin{remark}[Signed externality for net rather than normalised gross payoffs]\label{rem:netexternality}
For $R_i(s)>0$, the cross-effect on the debtor-side payoff is
\begin{equation}\label{eq:netcross}
\frac{\partial U_i^{\mathrm{net}}}{\partial s_j}
=\chi_i L_{ji}\left[\bar G_i(R_i/\kappa_i)
+\psi_i\EE\bigl(Y_i\ind_{\{\kappa_iY_i>R_i\}}\bigr)\right]\ge0.
\end{equation}
At $R_i=0<e_i+z_i$, the payoff formulas apply to one-sided variations that keep the bank's own action feasible. Adding interest from bank $j$ changes the fixed-own-payment cross-effect by $-k_j L_{ji}$.
\end{remark}

\section{Network Equilibrium and Latent Fragility}\label{sec:equilibrium}
\subsection{The repayment operator and general equilibrium results}
Write $D_{\pb,\A}=\operatorname{diag}(\pb_i)_{i\in\A}$ and let $I$ denote the identity matrix of the appropriate dimension. On the active debtor set, define the unamplified exposure matrix, its confidence-amplified counterpart, and the intercept vector:
\begin{equation}\label{eq:matrices}
M_0=D_{\pb,\A}^{-1}L_{\A\A}^{\top},\qquad
M=(I+D_\nu)M_0,\qquad
g_i=\frac{e_i-h_i-\nu_i\zb_i}{\pb_i},
\end{equation}
where $D_\nu=\operatorname{diag}(\nu_i)$. Explicitly,
$M_{ij}=(1+\nu_i)L_{ji}/\pb_i$ for $i,j\in\A$: an entry is positive exactly when debtor $j$ pays creditor $i$. The matrices depend on $L$ through both bilateral claims and outgoing totals; $g$ additionally depends on $(e,h,\nu)$. Thus
\begin{equation}\label{eq:system}
B(s)=\clip(g+Ms),\qquad s^*=B(s^*).
\end{equation}
Clipping is componentwise. In fixed-point expressions, $0$ and $\one$ are the zero and all-ones vectors on $\A$. Write $\rho(\cdot)$ for the spectral radius. The shock law enters the clearing operator through its reserve quantile $c_i$. For every law in Assumption~\ref{ass:scale}, the same box-constrained mixed complementarity conditions in Appendix~\ref{app:proofs} characterise equilibrium. Each fixed point is a generalised Nash equilibrium because $B_i$ is the unique maximiser over its counterparty-dependent feasible set.

\begin{proposition}[Existence, extremal equilibria, and uniqueness]\label{prop:lattice}
The operator $B:[0,1]^{|\A|}\to[0,1]^{|\A|}$ is continuous and nondecreasing (isotone). Its nonempty fixed-point set is a complete lattice with least equilibrium $s^-$ and greatest equilibrium $s^+$. Iteration from $0$ increases to $s^-$; iteration from $\one$ decreases to $s^+$. If
\begin{equation}\label{eq:rho}
\rho(M)<1,
\end{equation}
then the equilibrium is unique and best-response iteration converges geometrically from every initial profile. A sufficient, but generally more conservative, condition is $(1+\max_i\nu_i)\rho(M_0)<1$.
\end{proposition}
The mechanical matrix $M_0$ is similar to $\Pi_{\A\A}^{\top}$. If every active debtor has a directed chain of obligations to a pure creditor, the active relative-liability block is transient, and $\rho(M_0)<1$. This gives one sufficient condition for unique mechanical clearing. EN existence and regularity-based uniqueness do not require contraction; Appendix~\ref{app:en} separates these results. The strategic matrix can nevertheless have spectral radius at or above one because its rows are multiplied by $1+\nu_i$.

Fix an equilibrium partition $\A=\mathcal Z\cup\mathcal I\cup\mathcal F$ into zero, interior and full-payment banks. Write $D_{\pb,\mathcal I}=\operatorname{diag}(\pb_i)_{i\in\mathcal I}$ and let $\one_{\mathcal F}$ be the vector of ones indexed by $\mathcal F$. The interior coordinates satisfy
\begin{equation}\label{eq:regimes}
(I-M_{\mathcal I\mathcal I})s_{\mathcal I}=g_{\mathcal I}+M_{\mathcal I\mathcal F}\one_{\mathcal F}.
\end{equation}
If $\rho(M_{\mathcal I\mathcal I})<1$, the inverse exists and, while the partition remains unchanged,
\begin{equation}\label{eq:multiplier}
\frac{\partial s_{\mathcal I}}{\partial e_{\mathcal I}}=(I-M_{\mathcal I\mathcal I})^{-1}D_{\pb,\mathcal I}^{-1}
=\left(\sum_{k=0}^{\infty}M_{\mathcal I\mathcal I}^k\right)D_{\pb,\mathcal I}^{-1}.
\end{equation}
In payment units, $M^{\mathrm{pay}}_{\mathcal I\mathcal I}=D_{\pb,\mathcal I}M_{\mathcal I\mathcal I}D_{\pb,\mathcal I}^{-1}$ has entries $(1+\nu_i)\Pi_{ji}$. The column sums of $(I-M^{\mathrm{pay}}_{\mathcal I\mathcal I})^{-1}$ give the marginal gain in gross early payments per pound of opening cash at each interior bank. Its series sums the contributions of directed payment walks within the fixed interior set.

\begin{proposition}[Strongly connected localisation of possible multiplicity]\label{prop:scc}
Let $\mathcal C_1,\ldots,\mathcal C_K$ be the strongly connected components (SCCs) of the directed active-bank payment-dependency graph, with an edge $j\to i$ when $L_{ji}>0$. Then
\begin{equation}\label{eq:scc}
\rho(M)=\max_{1\le k\le K}\rho(M_{\mathcal C_k\mathcal C_k}).
\end{equation}
If the strategic game has more than one equilibrium, at least one SCC satisfies $\rho(M_{\mathcal C_k\mathcal C_k})\ge1$.
\end{proposition}
\begin{corollary}[Acyclic payment-dependency networks]\label{cor:acyclic}
If the active-bank payment-dependency graph is acyclic, its amplification matrix $M$ is nilpotent. The strategic equilibrium is therefore unique.
\end{corollary}

Strongly connected blocks locate the feedback that can sustain coordination failure, echoing the role of dependency cycles in \citet{JacksonPernoud2024}. The liquidity bounds in Proposition~\ref{prop:coordination} determine when full payment and an all-zero freeze coexist. Proofs are in Appendix~\ref{app:proofs}.

\subsection{A latent coordination region}
\begin{proposition}[Full payment and payment freezes]\label{prop:coordination}
For every active institution, let $h_i=\omega_ic_i$ and $\nu_i=\psi_ic_i$.
\begin{enumerate}[label=(\roman*),leftmargin=2.2em,itemsep=3pt]
\item Zero repayment $s=0$ is an equilibrium if and only if
\begin{equation}\label{eq:zeroeq}
e_i\le h_i+\nu_i\zb_i\qquad\text{for every }i\in\A.
\end{equation}
\item Full repayment $s=\one$ is an equilibrium if and only if
\begin{equation}\label{eq:fulleq}
e_i+\zb_i\ge\pb_i+h_i\qquad\text{for every }i\in\A.
\end{equation}
\item These equilibria coexist if and only if both conditions hold. Equivalently, for every active bank,
\begin{equation}\label{eq:band}
\pb_i-\zb_i+h_i\le e_i\le h_i+\nu_i\zb_i.
\end{equation}
The difference between the upper and lower endpoints is $(1+\nu_i)\zb_i-\pb_i$. Raising confidence sensitivity relaxes the freeze condition while leaving the full-repayment condition unchanged.
\end{enumerate}
\end{proposition}

If $e_i>0$ at every active bank, an all-zero equilibrium requires $c_i>0$ (equivalently $\delta_i<\chi_i$) at each bank; when $c_i=0$, $B_i(0)>0$. Coexistence also requires $(1+\nu_i)\zb_i\ge\pb_i$ bank by bank.

The set in \eqref{eq:band} can have nonempty interior even when $\rho(M_0)<1$. That observation implies a substantive separation between \emph{mechanical regularity} and \emph{strategic robustness}: leakage to outside creditors can ensure unique mechanical clearing without eliminating the possibility of a strategic freeze.

\paragraph{An all-zero fragility margin}
Define the system-level freeze margin from the bank-level intercepts in \eqref{eq:matrices}:
\begin{equation}\label{eq:fragilitymargin}
\Gamma_0:=\max_{i\in\A} g_i.
\end{equation}
The all-zero profile is an equilibrium exactly when $\Gamma_0\le0$. The intercept $g_i$ measures opening cash less the reserve target under zero incoming payments, scaled by obligations. The margin uses opening cash, bank-level payment totals and calibrated or estimated reserve parameters. It can flag a system-wide freeze that coexists with current full settlement. To assess partial freezes, $Z^-:=|\{i\in\A:s_i^-=0\}|/|\A|$ measures the share of active banks paying zero in the least equilibrium, and $RR(s^-)$ measures aggregate repayment. The extremal gap $\Xi:=\|s^+-s^-\|_\infty$ measures the spread between equilibria.

\begin{corollary}[Latent fragility under unique mechanical clearing]\label{cor:latent}
Suppose $\rho(M_0)<1$, $\A\ne\varnothing$, and $e_i+\zb_i\ge\pb_i+h_i$ for all active banks. Then $\one$ is the unique mechanical equilibrium and remains a strategic equilibrium as the sensitivities $\psi_i$ vary holding $h$ fixed. If in addition $e_i\le h_i+\nu_i\zb_i$ for every active bank, the strategic game has both $0$ and $\one$ as equilibria.
\end{corollary}
At full early settlement, $\Delta_i=0$, so the full-payment condition is independent of $\nu_i$. With $L,e,h$ fixed, the same observed payments can therefore be consistent with different reserve sensitivities and different sets of equilibria. We call the possibility of a freeze alongside full payment \emph{latent fragility}. The margin $\Gamma_0$ screens for an all-zero equilibrium. In Section~\ref{subsec:latentex}, the correspondent's opening cash gives $\Gamma_0=1.5$, even though a partial freeze delays \pounds233 million. Bilateral payment links help locate partial freezes and identify targets for intervention.

\begin{corollary}[Coexistence frontier across directed networks]\label{cor:homogeneous}
Suppose active banks share outgoing obligations $\bar p$, incoming claims $\bar z$, behavioural parameters and the funding-need law, with $\bar p>\bar z>0$. Write $m^Y=\EE[Y]$ and $\eta=\psi m^Y$. For some common pass-through $\eta\le\bar\eta$, where $\bar\eta>0$, a positive-width interval of common external liquidity $e$ supports both corner equilibria exactly when
\begin{equation}\label{eq:threshold_general}
\frac{c}{m^Y}>\frac{\bar p/\bar z-1}{\bar\eta}.
\end{equation}
The mechanical clearing vector is unique for every $e$.
\end{corollary}
Indeed, the coexistence band in \eqref{eq:band} has width $\bar z(1+\eta c/m^Y)-\bar p$, which is largest at $\eta=\bar\eta$. Also $M_0$ has constant row sum $\bar z/\bar p<1$, which suffices for mechanical uniqueness. The frontier depends on the reserve quantile relative to mean funding need, and applies across directed wirings with the same totals. The wiring still governs propagation and can affect other equilibria.

\begin{corollary}[A confidence threshold for the freeze equilibrium]\label{cor:threshold}
Fix $\bar\nu\ge0$ with $\bar\nu_i=0$ whenever $c_i=0$, and let $\nu_i(q)=q\bar\nu_i$, $q\ge0$. For each active bank with $\bar\nu_i\zb_i>0$, define $q_i^{\mathrm{conf}}=(e_i-h_i)/(\bar\nu_i\zb_i)$. If each active bank with $\bar\nu_i\zb_i=0$ satisfies $e_i\le h_i$, then the zero profile is an equilibrium for all
\begin{equation}\label{eq:threshold}
q\ge q^*:=\max\left\{0,\max_{i\in\A:\,\bar\nu_i\zb_i>0}q_i^{\mathrm{conf}}\right\}.
\end{equation}
If no active bank has a positive coefficient $\bar\nu_i\zb_i$, set $q^*=0$ under the stated inequalities.
If an active bank with $\bar\nu_i\zb_i=0$ has $e_i>h_i$, zero repayment is impossible for every $q$.
\end{corollary}

\subsection{Comparative statics and interventions}
For a fixed liability matrix and other primitives, higher $h_i$ or $\nu_i$ weakly reduces $B_i(s)$ pointwise because
\begin{equation}\label{eq:B_hnu}
B_i(s)=\clip\!\left(\frac{e_i+z_i-h_i-\nu_i\Delta_i}{\pb_i}\right),
\qquad \Delta_i\ge0.
\end{equation}
By isotonicity, matched greatest and least equilibria weakly decrease as $h_i$ or $\nu_i$ rises. Higher external resources $e_i$ have the reverse effect. A higher late rate weakly lowers $c_i$, desired reserves and confidence feedback. This comparison fixes private costs and unrestricted liquidity and holds the pledge sufficient. A cash injection changes the intercept of the repayment rule at fixed feedback coefficients. Raising $\chi_i$ weakly increases desired retention.

At the all-zero profile, a bank with $e_i>0$ optimally pays zero exactly when its late charge satisfies
\begin{equation}\label{eq:interestfreeze}
k_i\le k_i^{\mathrm{freeze}}:=
\chi_i\bar G_i\!\left(\frac{e_i}{\omega_i+\psi_i\zb_i}\right)-d_i^{\mathrm{priv}} .
\end{equation}
This is the condition that the right derivative of payoff at zero be nonpositive. A charge above the threshold for any such bank rules out the all-zero equilibrium. At that profile, banks with $e_i\le0$ must already pay zero because they lack date-0 cash.

The all-zero equilibrium disappears as soon as some bank has $e_i>h_i+\nu_i\zb_i$.

Let $u_j$ be the coordinate vector for the cash recipient $j$. For full payment by a group of active banks $C$, the required budget has infimum
\begin{equation}\label{eq:injectiontarget}
Q_j(C):=\inf\{q\ge0:\ s_i^-(e+qu_j)=1\text{ for all }i\in C\},
\qquad \inf\varnothing=\infty .
\end{equation}
The least equilibrium tests whether the target is met in every equilibrium. To determine attainment at a finite cutoff, evaluate the least equilibrium at that budget. If $j\notin C$ and has no directed payment path to $C$, the iteration from zero on $C$ and its ancestors is unaffected. A bank already at full payment in the least equilibrium cannot raise that equilibrium through an injection: only its own response changes, that response remains clipped at one at the old fixed point, and monotonicity gives the same least equilibrium. For a group still short of full payment, either test gives $Q_j(C)=\infty$. Equation~\eqref{eq:multiplier} gives marginal gains within a fixed, contractive repayment regime. The threshold in \eqref{eq:injectiontarget} allows the clipping partition to change.

\subsection{Equilibrium selection and adjustment}\label{subsec:selection}
For the symmetric two-bank specification $b(x)=\clip(g+mx)$, consider synchronous best-response updates of payment conjectures. These conjectures may lie outside the jointly cash-feasible set.
\begin{proposition}[Symmetric best-response threshold]\label{prop:basin}
Suppose $m>1$ and $1-m<g<0$. Then the two-bank system $B_1(s_2)=b(s_2)$, $B_2(s_1)=b(s_1)$ has corner equilibria $(0,0)$ and $(1,1)$ and an interior symmetric equilibrium $(x^*,x^*)$, where
\begin{equation}\label{eq:basinthreshold}
x^*=-\frac{g}{m-1}\in(0,1).
\end{equation}
Under synchronous best-response adjustment starting on the symmetric diagonal $s_1^{(0)}=s_2^{(0)}=x$, iteration converges to $(0,0)$ for $x<x^*$, remains at $(x^*,x^*)$ for $x=x^*$, and converges to $(1,1)$ for $x>x^*$. Both corner equilibria are locally stable to sufficiently small perturbations in the full two-dimensional strategy space.
\end{proposition}

The threshold characterises symmetric conjectures under this adjustment rule. Asymmetric conjectures can alternate.

\section{Systemic Consequences and the Mechanical Benchmark}\label{sec:systemic}
\subsection{Mechanical-versus-strategic repayments}
Let $\Phi:[0,1]^{|\A|}\to[0,1]^{|\A|}$ be the EN-type maximal-feasible-repayment operator from Definition~\ref{def:mechanical}.

\begin{proposition}[Mechanical clearing bounds strategic repayment]\label{prop:mechanical}
For every profile $s$, $B(s)\le\Phi(s)$ componentwise. Let $s^+,s^-$ be the greatest and least strategic equilibria, and let $s_\Phi^+,s_\Phi^-$ be the corresponding extremal mechanical fixed points. Then
\begin{equation}\label{eq:mechanicalorder}
s^+\le s_\Phi^+,\qquad s^-\le s_\Phi^-.
\end{equation}
If $\delta_i\ge\chi_i$ for every active institution, $B\equiv\Phi$ and the two equilibrium sets coincide.
\end{proposition}
A bank's early shortfall can combine insufficient cash with deliberate withholding of cash that remains payable. Lower payments also reduce the cash available to counterparties.

\subsection{An exact decomposition of capacity and strategic withholding}
At a strategic equilibrium $s^\G$, let $p_i^\G=\pb_i s_i^\G$, $a_i^\G=a_i(s^\G_{-i})$, and $q_i^\G=\min\{\pb_i,a_i^\G\}$ be the monetary amount that bank $i$ could currently repay. Define the capacity shortfall $\mathrm{Cap}_i^\G$ and directly withheld payable resources $W_i^\G$ by
\begin{equation}\label{eq:two_losses}
\mathrm{Cap}_i^\G=(\pb_i-a_i^\G)^+,
\qquad W_i^\G=q_i^\G-p_i^\G=w_i(s^\G_{-i})\ge0.
\end{equation}
Then the contractual shortfall decomposes \emph{exactly} into
\begin{equation}\label{eq:bankdecomp}
\pb_i-p_i^\G
=\underbrace{(\pb_i-a_i^\G)^+}_{\text{current capacity shortfall}}
+\underbrace{\left[\min\{\pb_i,a_i^\G\}-p_i^\G\right]}_{\text{strategic withholding of payable resources}}.
\end{equation}

This includes withholding by banks that could repay only part of their obligations. Since equilibrium inflows already reflect counterparties' choices, the capacity term also incorporates network effects; the decomposition is evaluated at the strategic outcome.

Let $\pbtot:=\sum_{i\in\A}\pb_i>0$. Define the aggregate repayment and direct-withholding rates by
\begin{equation}\label{eq:RR}
RR(s):=\frac{\sum_{i\in\A}\pb_i s_i}{\pbtot},\qquad
\mathrm{SWR}(s^\G):=\frac{\sum_{i\in\A}W_i^\G}{\pbtot}.
\end{equation}
Take $(s^\G,s^{\EN})=(s^+,s_\Phi^+)$ or $(s^-,s_\Phi^-)$ to match the extremal selections. The \emph{excess strategic shortfall} in early payments against the mechanical benchmark is
\begin{equation}\label{eq:ESR}
ESR:=RR(s^{\EN})-RR(s^\G)\ge0.
\end{equation}
Under matched selections, $s^{\EN}\ge s^\G$ so $s^{\EN}=\Phi(s^{\EN})\ge\Phi(s^\G)$. Evaluating maximum feasible payments at the strategic inflows separates the two sources of the aggregate gap:
\begin{equation}\label{eq:ESRcapacity}
ESR=\underbrace{\mathrm{SWR}(s^\G)}_{\text{direct withholding}}
+\underbrace{RR(s^{\EN})-RR(\Phi(s^\G))}_{\text{network-induced capacity loss}} .
\end{equation}
Both terms are nonnegative. The first measures cash that banks could pay at their actual inflows and choose to retain. The second measures how much payment capacity disappears as counterparties pay less. Setting all $\psi_i$ to zero at fixed remaining parameters gives a second diagnostic. If $s^{\G,0}$ denotes the matching extremal equilibrium of that counterfactual, then
\begin{equation}\label{eq:ESRdecomp}
ESR=\underbrace{RR(s^{\EN})-RR(s^{\G,0})}_{\text{constant-stress precautionary shortfall}}
+\underbrace{RR(s^{\G,0})-RR(s^\G)}_{\text{increment from confidence-dependent stress}}.
\end{equation}
This decomposition shows how much of the gap arises from precautionary retention at constant funding risk and how much extra confidence-sensitive funding needs add. Both terms are nonnegative under matched extremal selections.

\subsection{Late settlements versus eventual creditor losses}
The shortfall in early principal comprises late principal payments and ultimate principal losses. For the continuation-cash vector $X=(X_i)_{i\in\A}$, define the two fractions of original principal by
\begin{equation}\label{eq:terminalrates}
\mathrm{Late}(s,X)=\frac{\sum_{i\in\A}Q_i^1(s,X_i)}{\pbtot},\qquad
\mathrm{Loss}(s,X)=\frac{\sum_{i\in\A}\mathcal D_i^1(s,X_i)}{\pbtot}.
\end{equation}
Proposition~\ref{prop:terminal} immediately yields the accounting identity
\begin{equation}\label{eq:terminaldecomposition}
1-RR(s)=\mathrm{Late}(s,X)+\mathrm{Loss}(s,X).
\end{equation}
Late interest is an additional transfer from pledged balances. Counterparties' delays can reduce other banks' available cash and increase their principal defaults, even though each bank's own principal default is invariant to its initial action at a fixed counterparty profile.

\begin{proposition}[Fixed-budget accounting comparison of terminal losses]\label{prop:terminalloss}
Hold the nominal obligations $L$ and external date-0 resources $e$ fixed. Assume (i) the same exogenous nonnegative continuation-cash vector $X$ in both counterfactuals, independent of the settlement profile; (ii) interim bridge funding does not permanently consume reserves available for terminal principal; and (iii) late payments received from other banks are not recycled into the same-round date-1 settlement budgets. Under the terminal rule in \eqref{eq:terminalrule}, if $s'\ge s''$ componentwise, then
\begin{equation}\label{eq:terminalorder}
\mathcal D_i^1(s',X_i)\le\mathcal D_i^1(s'',X_i)
\quad\text{for every active }i.
\end{equation}
Consequently, for the same greatest- or least-equilibrium selection rule in the mechanical and strategic models,
\begin{equation}\label{eq:terminalorderEN}
\mathrm{Loss}(s^{\G},X)\ge\mathrm{Loss}(s^{\EN},X).
\end{equation}
Ample continuation cash can leave terminal principal losses at zero under both rules. Terminal default records can therefore miss a morning payment freeze.
\end{proposition}

The ordering holds pointwise in $X$ and under a common continuation-cash distribution. It follows from $\mathcal D_i^1(s,X_i)=[\pb_i-a_i(s_{-i})-X_i]^+$ and the monotonicity of incoming cash. Simultaneous recycling of terminal receipts would require solving the continuation clearing problem.

\subsection{Private payoff comparisons}

Across equilibria, $U_i^{\mathrm{net}}$ accounts for the state-dependent funding cost that is normalised out of $U_i$.

\begin{proposition}[Debtor-side payoffs at ordered repayment equilibria]\label{prop:pareto}
If $s'$ and $s''$ are strategic equilibria with $s'\ge s''$ componentwise, then $U_i^{\mathrm{net}}(s')\ge U_i^{\mathrm{net}}(s'')$ for every active institution. In particular, each bank has a weakly higher debtor-side payoff at the greatest equilibrium than at any other equilibrium. If the distribution of $X_i$ is invariant across other banks' payment profiles, the same ordering holds for the completed payoffs $\mathcal J_i$ in \eqref{eq:completepayoff} whenever $\mathrm{Pen}_i^{\mathrm{def}}$ is nondecreasing.
\end{proposition}

Higher counterparty payments expand feasible settlement and reduce expected interim shortages. With profile-invariant continuation cash they also weakly reduce default penalties. The result ranks debtor-side payoffs, which include interest paid. Incoming late interest falls as counterparties pay earlier.

\section{Analytical Illustrations}\label{sec:illustrations}
\subsection{A margin-window freeze with full final repayment}\label{subsec:latentex}
Two clearing dealers have wholesale payments due at 09:00. A cash-rich correspondent pays Dealer 1, and a client bank relies on Dealer 2's payment to meet a separate obligation. Each dealer chooses its payment based on what it expects to receive and the cash it may need for a collateral call later that day. A missed morning receipt reduces cash on hand and can increase later cash needs if other lenders cut short-term lines or demand more collateral. A temporary cash bridge unwinds before the next morning's cut-off, when deferred principal and late interest fall due. This timing is motivated by the margin-related liquidity demands documented in March 2020 \citep{BCBSCPMIIOSCO2022}. The network and balances are illustrative. Amounts are in millions of pounds.

Order the active banks as Dealer 1, Dealer 2, correspondent and client bank. The first matrix gives wholesale obligations, with debtors in rows. The last vector gives obligations to an outside creditor:
\begin{equation}\label{eq:latentL}
L_{\A\A}=\begin{pmatrix}
0&100&0&0\\
95&0&0&10\\
10&0&0&0\\
0&0&0&0
\end{pmatrix},\qquad
L_{\A,\mathrm{out}}=(15,5,0,20)^\top .
\end{equation}
The four internal links give $\pb=(115,110,10,20)$ and $\zb=(105,100,0,10)$. Unrestricted cash is $e=(22,32,15,12)$. Mechanical clearing has feedback radius $\sqrt{(95/115)(100/110)}\simeq0.867$ and uniquely pays all \pounds255 million at the morning cut-off.

\paragraph{Funding and deferral costs}
Use the September 2019 US Treasury repo 99th percentile of 9\% \citep{AfonsoEtAl2021} and the contemporaneous 2.10\% interest rate on reserves \citep{FederalReserve2019} to benchmark a one-day emergency funding premium $\chi=(0.09-0.021)/360$, or 1.92 basis points per uncovered pound. All banks face one basis point of contractual late interest, $k=0.00010$, backed by pledged cash $C_i=0.00020\pb_i$. The dealers' total deferral cost is $\delta=\chi/1.44$. Use GPD shape $\xi=0.25$ with scales $\omega_1=20$ and $\omega_2=30$. These imply expected interim cash needs of 26.67 and 40 when receipts arrive in full. Set $\psi_1=\psi_2=0.75$, so each missing pound adds one pound to expected interim need. Both dealers have $c\simeq0.3817805$ and $\nu=0.75c\simeq0.2863353$, with distinct baseline reserve targets $h_i=\omega_i c$. For the correspondent and client bank set $\omega_i=10$, $\psi_i=0$ and $\delta_i=\chi$, so they pay as much as their cash allows.

\paragraph{Three equilibria under the same cash positions}
The correspondent always pays 10. Hence $z_1=10+95s_2$ and $z_2=100s_1$. The client bank remits $\min\{20,12+10s_2\}$ to its outside creditor. With $b_i(x)=\clip(\gamma_i+\beta_ix)$, the dealers' best responses are
\begin{equation}\label{eq:latentBR}
\begin{aligned}
b_1(s_2)&=\clip\!\left(\frac{32-h_1-95\nu+95(1+\nu)s_2}{115}\right),\\
b_2(s_1)&=\clip\!\left(\frac{32-h_2-100\nu+100(1+\nu)s_1}{110}\right).
\end{aligned}
\end{equation}
In the low equilibrium each dealer has 32 available but holds it all: the desired reserves are 34.84 and 40.09. The lost payment from Dealer 2 leaves the client bank able to meet only 12 of its 20 due. In the high equilibrium both dealers pay in full and retain 12 and 22 respectively, above their reserve targets of 7.64 and 11.45. The client bank receives 10 and pays in full. When $\psi_i=0$ for both dealers, the baseline reserve targets still apply and full payment is unique by contraction.

\begin{table}[htbp]\centering\small
\caption{Early settlement in the margin-window network}\label{tab:marginoutcomes}
\begin{tabular}{@{}lrrrrr@{}}\toprule
Equilibrium & Dealer 1 & Dealer 2 & Client & \shortstack{Early payments\\(\pounds m)} & \shortstack{Late interest\\(\pounds000)}\\\midrule
Dealer freeze & 0.000 & 0.000 & 0.600 & 22.00 & 23.30\\
Intermediate & 0.424 & 0.422 & 0.811 & 121.35 & 13.36\\
Full payment & 1.000 & 1.000 & 1.000 & 255.00 & 0.00\\\bottomrule
\end{tabular}
\par\vspace{3pt}{\footnotesize\emph{Notes.} The first three numeric columns are repayment fractions. The correspondent pays in full in each equilibrium. Late interest is $k$ times aggregate deferred principal.}
\end{table}

Set continuation cash at $X=(100,80,0,10)$ from separately scheduled receipts and cash buffers released at the final cut-off. Then $e_i+X_i>\pb_i$ for every bank, and each pledge covers its late charge. All principal and interest therefore settle by date 1 even after the morning freeze. The freeze delays \pounds233 million, just over 91\% of scheduled morning payments.

\paragraph{Conditional adjustment and a targeted late charge}
With the correspondent paying in full, Dealer 1's fraction follows $T=b_1\circ b_2$ every two steps under synchronous best responses. The intermediate equilibrium has Dealer 1 at $x^*\simeq0.423673$, where $T$ crosses the diagonal with slope $\beta_1\beta_2\simeq1.243$. Initial Dealer 1 conjectures below $x^*$ converge to zero on this subsequence and those above it converge to one. Dealer 2 has threshold $0.421924$ under $b_2\circ b_1$. Starting both dealer conjectures below their thresholds leads to the freeze, and starting both above leads to full payment.

Hold private delay costs and pledges fixed and increase the charge $k_1$ of Dealer 1 only. The exact zero-payments outcome for the core disappears above about 1.038 basis points. Full early settlement becomes unique once the charge strictly exceeds the threshold of approximately 1.143 basis points, up to the two-basis-point pledge limit. At the higher threshold, Dealer 1's best response to zero payment reaches the point where any further increase makes Dealer 2 pay a positive amount. Appendix~\ref{app:numerics} derives the cut-offs and Section~\ref{subsec:arrears} compares this policy with early access to unsettled receipts.

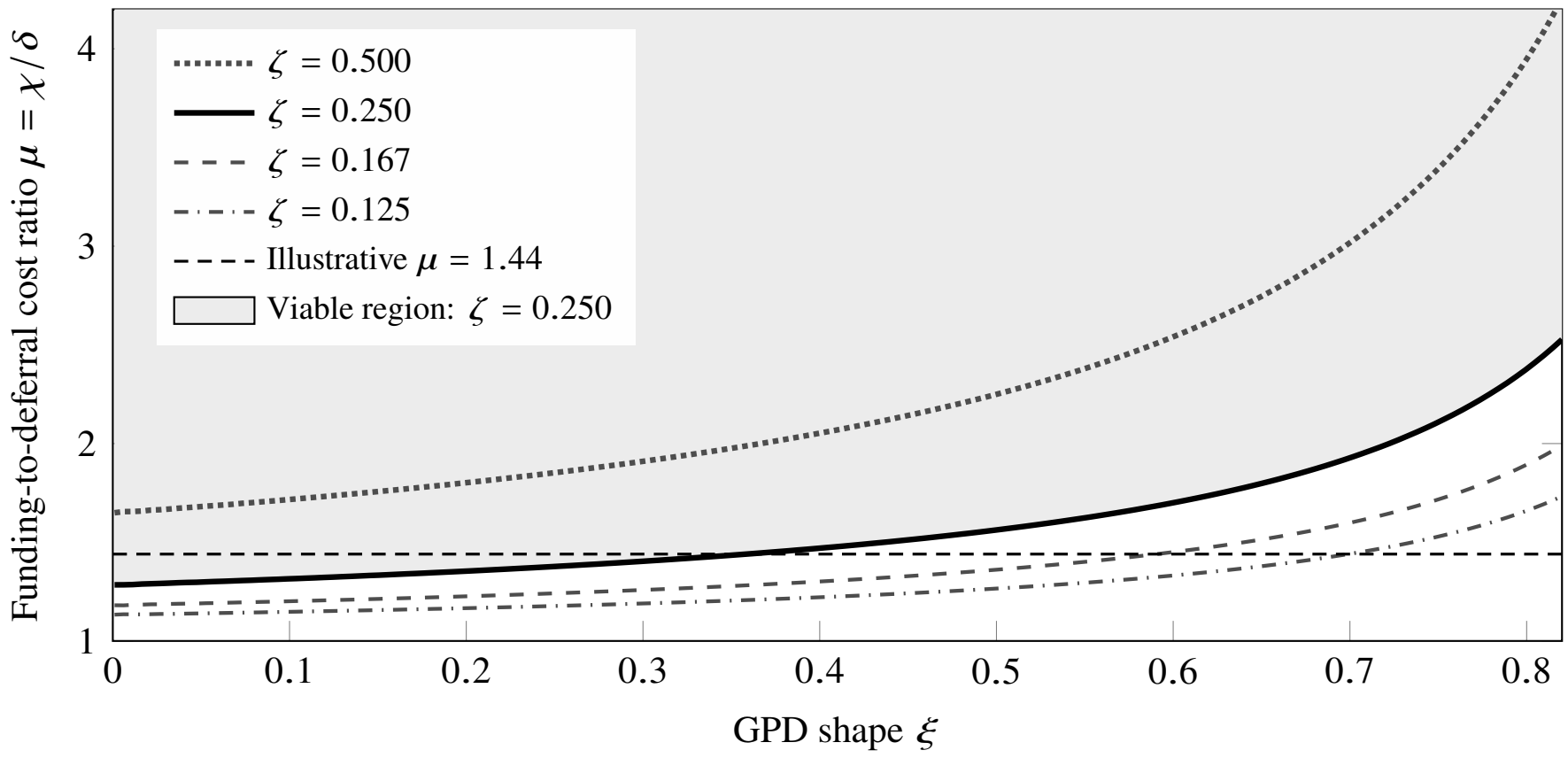
\begin{figure}[htbp]\centering
\begin{tikzpicture}
\begin{axis}[
 width=0.92\linewidth,height=6.3cm,xmin=0,xmax=0.82,ymin=1.0,ymax=4.2,
 xlabel={GPD shape $\xi$},ylabel={Funding-to-deferral cost ratio $\mu=\chi/\delta$},
 ticklabel style={font=\footnotesize},every axis label/.append style={font=\footnotesize},
 legend style={font=\scriptsize,at={(0.03,0.97)},anchor=north west,
   fill=white,fill opacity=0.96,text opacity=1,draw=none},
 legend cell align=left]
\addplot[name path=capone,forget plot,draw=none,domain=0.001:0.82,samples=160]
 {pow(1+0.25*x/(1-x),1/x)};
\addplot[name path=topbound,forget plot,draw=none,domain=0.001:0.82] {4.2};
\addplot[forget plot,fill=gray!15,draw=none] fill between[of=capone and topbound];
\addplot[forget plot,black!70,densely dotted,very thick,domain=0.001:0.82,samples=160]
 {pow(1+0.5*x/(1-x),1/x)};
\addplot[forget plot,black,solid,very thick,domain=0.001:0.82,samples=160]
 {pow(1+0.25*x/(1-x),1/x)};
\addplot[forget plot,black!70,dashed,thick,domain=0.001:0.82,samples=160]
 {pow(1+(1/6)*x/(1-x),1/x)};
\addplot[forget plot,black!70,dashdotted,thick,domain=0.001:0.82,samples=160]
 {pow(1+0.125*x/(1-x),1/x)};
\addplot[forget plot,black,densely dashed,semithick,domain=0:0.82] {1.44};
\addlegendimage{black!70,densely dotted,very thick}\addlegendentry{$\zeta=0.500$}
\addlegendimage{black,solid,very thick}\addlegendentry{$\zeta=0.250$}
\addlegendimage{black!70,dashed,thick}\addlegendentry{$\zeta=0.167$}
\addlegendimage{black!70,dashdotted,thick}\addlegendentry{$\zeta=0.125$}
\addlegendimage{black,densely dashed,semithick}\addlegendentry{Illustrative $\mu=1.44$}
\addlegendimage{area legend,fill=gray!15,draw=none}\addlegendentry{Viable region: $\zeta=0.250$}
\end{axis}
\end{tikzpicture}
\caption{Coexistence over a positive-width range of external liquidity in homogeneous directed networks with GPD funding needs. The condition is $\mu>\mu_{\min}(\xi;\zeta)$, where $\zeta=(\bar p/\bar z-1)/\bar\eta$. At $\bar p/\bar z=1.25$, the curves correspond to caps $\bar\eta=0.5,1,1.5,2$. Shading marks the viable region for $\zeta=0.25$. The horizontal dashed line gives $\mu=1.44$.}\label{fig:muxi}
\end{figure}

\paragraph{Cost-ratio and tail-shape frontier}
To compare the range of cash buffers supporting both corner equilibria across wirings, consider the homogeneous benchmark of Corollary~\ref{cor:homogeneous}. For $\mu=\chi/\delta>1$, let $c(\mu,\xi)=(\mu^\xi-1)/\xi$ with $c(\mu,0)=\log\mu$. At maximum pass-through $\bar\eta>0$ and $0<\xi<1$, the positive-width region is
\begin{equation}\label{eq:capfrontier}
(1-\xi)c(\mu,\xi)>\zeta
\quad\Longleftrightarrow\quad
\mu>\mu_{\min}(\xi;\zeta)
=\left(1+\frac{\zeta\xi}{1-\xi}\right)^{1/\xi},
\qquad
\zeta=\frac{\bar p/\bar z-1}{\bar\eta}.
\end{equation}
At $\xi=0$, $\mu_{\min}(0;\zeta)=e^\zeta$. Figure~\ref{fig:muxi} shows the frontier. The baseline scale $\omega$ shifts the admissible cash range through $h=\omega c$ without affecting its width. With $\bar p=100$, $\bar z=80$, $\omega=20$, $\mu=1.44$ and unit pass-through, the interval at $\xi=0.25$ is $[27.636,30.542]$. Table~\ref{tab:capchecks} shows how the viable tail shape and interval width change with the pass-through cap.

\begin{table}[htbp]\centering\small
\caption{Sensitivity of coexistence to the maximum expected funding-need pass-through}\label{tab:capchecks}
\begin{tabular}{@{}rrrr@{}}\toprule
Cap $\bar\eta$ & Upper cutoff for $\xi$ at $\mu=1.44$ & Largest $e$-band at $\xi=0.25$ & Coexistence at $\xi=0.25$ \\\midrule
$0.50$ & None & --- & No \\
$0.75$ & $0.1029$ & --- & No \\
$1.00$ & $0.3582$ & $2.907$ & Yes \\
$1.50$ & $0.5904$ & $14.360$ & Yes \\
$2.00$ & $0.6990$ & $25.814$ & Yes \\\bottomrule
\end{tabular}
\par\vspace{3pt}{\footnotesize\emph{Notes:} Network ratio $\bar p/\bar z=1.25$, $\mu=1.44$, and baseline $\omega=20$. The maximum $e$-band is $[80\bar\eta(1-\xi)c(\mu,\xi)-20]^+$; dashes mean the nondegenerate coexistence band is empty. Each row imposes the indicated pass-through cap.}
\end{table}

\paragraph{Rewiring at fixed payment totals}\label{par:rewiring}
Four active banks each owe 90 to active peers and 10 to an outside creditor and are each owed 90 by peers. Set $e=19$ and $\eta=0.5$ (so $\psi=0.375$) with $\omega=20$, $\xi=0.25$ and $\mu=1.44$ from the homogeneous benchmark. In the diversified network, each owes 30 to each other bank. In the concentrated network, the internal liability matrix is
\[
L^{\mathrm{conc}}_{\A\A}
=\begin{pmatrix}
0&89.5&0.5&0\\
89.5&0&0&0.5\\
0&0.5&0&89.5\\
0.5&0&89.5&0
\end{pmatrix}.
\]
The directed links connect two tightly linked pairs. Corollary~\ref{cor:homogeneous} gives the same coexistence interval $[17.636,20.521]$ and $\rho(M_0)=0.9$ for both designs. Both have unique mechanical full payment at $e=19$. The uniform response and its threshold $x^*\simeq0.5271$ are identical. From the same starting conjecture, synchronous best responses converge to a complete freeze in the diversified network and a freeze in one pair in the concentrated network (Table~\ref{tab:topology}). In the concentrated network, a pair's threshold is $0.6573$ if the other pair stays at zero and $0.4103$ if it stays at one. Appendix~\ref{app:numerics} verifies these equilibria and adjustment paths.

\begin{table}[H]\centering\small
\caption{Payment links at fixed bank-level totals}\label{tab:topology}
\begin{tabular}{@{}lll@{}}\toprule
& Diversified & Concentrated\\\midrule
Number of equilibria & 3, all uniform & 9, including 6 nonuniform\\
Examples of additional equilibria & --- & $(1,1,0,0)$ and $(0,0,0.6573,0.6573)$\\
First pair's first update & $0.4786$ & $0.6215$\\
Second pair's first update & $0.5507$ & $0.4078$\\
Limit & $(0,0,0,0)$ & $(1,1,0,0)$\\\bottomrule
\end{tabular}
\par\vspace{3pt}{\footnotesize\emph{Notes:} The adjustment experiment starts from the same cash-feasible conjecture $(0.62,0.62,0.41,0.41)$ in both networks and iterates synchronous best responses. Fractions are rounded.}
\end{table}

\subsection{Payment amplification in a unique equilibrium}\label{subsec:nonmultiple}
The sterling overnight market exhibits a core of highly connected banks and asymmetric borrowing and lending relationships \citep{WetheriltEtAl2010}. German bilateral data also show tiering around money-centre intermediaries \citep{CraigVonPeter2014}. We illustrate these patterns with two dealers and four peripheral banks. There are ten directed links among the active banks out of thirty possible. The dealers owe each other different amounts, and peripheral banks trade through their respective dealers. Link amounts measure cash due at the early cut-off in millions of pounds.

The liabilities among active banks and the amounts due to an outside creditor are given in debtor-row order by
\begin{equation}\label{eq:ex2balance}
L_{\A\A}=\begin{pmatrix}
0&90&0&0&14&10\\
74&0&35&27&0&0\\
0&56&0&0&0&0\\
0&34&0&0&0&0\\
27&0&0&0&0&0\\
18&0&0&0&0&0
\end{pmatrix},\qquad
L_{\A,\mathrm{out}}=(46,44,24,21,18,12)^\top .
\end{equation}
Set early cash at $e=(100,18,32,22,30,20)$ and use GPD shape $\xi=1/2$. The scale vector is $\omega=(40,36,41,20,15,15)$, with $\psi=(0,0.35,0.50,0.25,0,0)$. With the same bridge cost $\chi$ and late charge $k$ as in Section~\ref{subsec:latentex}, set $\delta_i=\chi/(1+c_i/2)^2$ for $c=(0,0.2,0.4,0.1,0,0)$. This yields positive private delay costs. Give each bank an interest pledge equal to $k\pb_i$. Mechanical and strategic feedback radii are $0.714$ and $0.755$, so each clearing rule has a unique equilibrium.

\begin{table}[H]\centering\small
\caption{A dealer--client funding squeeze with unique early repayment}\label{tab:nonmultiple}
\begin{tabular}{@{}lrrrr@{}}\toprule
Bank & Mechanical & \shortstack{Feasible at\\strategic inflows} & Strategic & Withheld\\\midrule
1: money-centre dealer & 160.00 & 160.00 & 160.00 & 0.00\\
2: intermediary dealer & 180.00 & 166.58 & 157.18 & 9.40\\
3: regional bank & 67.00 & 62.56 & 45.27 & 17.29\\
4: regional bank & 49.00 & 45.58 & 43.49 & 2.09\\
5: peripheral bank & 44.00 & 44.00 & 44.00 & 0.00\\
6: peripheral bank & 30.00 & 30.00 & 30.00 & 0.00\\\midrule
Total & 530.00 & 508.72 & 479.94 & 28.77\\\bottomrule
\end{tabular}
\par\vspace{3pt}{\footnotesize\emph{Notes:} Amounts are in millions of pounds. Feasible at strategic inflows is $\pb_i\Phi_i(s^\G)$. Totals use unrounded values.}
\end{table}

Dealer 1 pays in full. Dealer 2 can pay in full under mechanical clearing, but its receipt from bank 3 falls from 46.90 to 31.69 in the strategic equilibrium. Bank 4 also pays less. Dealer 2 now has capacity to pay 166.58 of its 180 obligations and optimally withholds a further 9.40. Its payment to bank 3 falls from 35 to 30.56, feeding the shortage back to that bank. The early-payment shortfall rises from 3.64\% under mechanical clearing to 12.74\% in the unique strategic equilibrium. The 9.10-point gap comprises 5.23 points of direct withholding and 3.87 points of lost capacity from missed receipts, as in \eqref{eq:ESRcapacity}.

With $X_i=170$ for each active bank, $e_i+X_i>\pb_i$ throughout the network. The separate pledges fund late interest. All principal and interest ultimately settle, and the same early amplification operates at fixed asset prices. For small cash additions within the reported repayment regime, \eqref{eq:multiplier} gives increases of 1.917, 2.436, 2.268 and 1.000 in gross early payments per pound injected at banks 2, 3, 4 and 5, respectively. A marginal pound at regional bank 3 has the largest effect because it reaches Dealer 2 and feeds back through their payment links.

\subsection{Early access to arrears and settlement timing}\label{subsec:arrears}

Suppose an outside facility buys a fraction $\phi_i\in[0,1]$ of bank $i$'s outstanding incoming principal at par after date 0 and before the funding need. The cash proceeds $\phi_i\Delta_i(s_{-i})$ provide interim liquidity and are included once in $X_i$. The facility receives final principal payments on the assigned portion. The modified interim payoff is

\begin{equation}\label{eq:arrearsU}
U_i^{\mathrm{net,rec}}(s_i,s_{-i})=-\chi_i\EE\big[(\kappa_iY_i-R_i-\phi_i\Delta_i)^+\mid s_{-i}\big]
-\delta_i\pb_i(1-s_i).
\end{equation}

The assignment preserves the debtor's total principal due at date 1 and the late-interest charge on its date-0 arrears.

\paragraph{The simultaneous-payment counterexample}
Suppose that the entire deferred principal $A_i^1$ must be paid at exactly the same instant as the liquidity need, and that retained cash has no earlier service value. The net amount available to cover that need is
\begin{equation}\label{eq:ownarrears}
T_i(s_{-i})=R_i(s)+\phi_i\Delta_i(s_{-i})-A_i^1(s_i)
=a_i(s_{-i})-\pb_i+\phi_i\Delta_i(s_{-i}),
\end{equation}
independent of $s_i$. Under an objective $-\chi_i\EE[(\Lambda_i-T_i)^+]-\delta_i A_i^1$ with $\delta_i>0$, the bank strictly prefers the largest feasible date-0 payment. The strategic wedge comes from the use of retained cash before deferred principal falls due.

\begin{proposition}[Counterparty arrears recovery and effective amplification]\label{prop:arrears}
Under the scale-family assumptions and the deterministic recovery convention \eqref{eq:arrearsU}, define
\begin{equation}\label{eq:arrearsH}
H_i^{\mathrm{rec}}(s_{-i})=\big[h_i+(\nu_i-\phi_i)\Delta_i(s_{-i})\big]^+.
\end{equation}
The bank's unique best response is
\begin{equation}\label{eq:arrearsBR}
B_i^{\mathrm{rec}}(s_{-i})=\clip\left(\frac{e_i+z_i-H_i^{\mathrm{rec}}(s_{-i})}{\pb_i}\right).
\end{equation}
If $0\le \phi_i\le\nu_i$, the positive part is redundant, and the operator is globally clipped affine, obtained by replacing $\nu_i$ with $\nu_i-\phi_i$. Wherever bank $i$'s best response is interior and $H_i^{\mathrm{rec}}>0$,
\begin{equation}\label{eq:arrearsSlope}
\frac{\partial B_i^{\mathrm{rec}}}{\partial s_j}=\frac{(1+\nu_i-\phi_i)L_{ji}}{\pb_i}.
\end{equation}
For all $\phi_i\in[0,1]$, the operator remains continuous and isotone, though it can be only piecewise clipped affine when $\phi_i>\nu_i$.
\end{proposition}

At-par sale proceeds arrive before the interim funding need, while the stress scale still responds to missed date-0 payments. The proceeds raise the intercept and reduce the interior payment pass-through to $1+\nu_i-\phi_i$ while the desired reserve is positive.

Apply a common early recovery fraction $\phi$ to the two clearing dealers in Section~\ref{subsec:latentex}. The correspondent pays in full, so the missing amounts are $95(1-s_2)$ and $100(1-s_1)$. Substituting $\nu-\phi$ for $\nu$ in \eqref{eq:latentBR} eliminates zero repayment by both dealers above a recovery fraction of about $0.029868$. A low-payment equilibrium remains until recovery strictly exceeds the second threshold, about $0.055153$. At 6\% recovery full early settlement is unique even though the feedback radius is $1.063$. The radius drops below one only above $\phi\simeq0.132395$ in the affine range. Appendix~\ref{app:numerics} gives both threshold calculations.

\subsection{Cash targeting and claim prefunding}\label{subsec:targeting}
\paragraph{Which bank should receive cash?}
Return to the four-bank margin network and target full payment by the two dealers in every equilibrium. At the freeze, a marginal pound at the client raises gross morning payments by one pound. Both dealers' payments are locally flat at zero, and the correspondent already pays in full. At a budget of \pounds10.25 million, Dealer 1 alone restores full payment (Table~\ref{tab:targetcash}). The client can pay more to its own creditor, which has no payment path back to the core. Appendix~\ref{app:numerics} derives the threshold when cash can be split freely across banks. The infimum is \pounds10.06730 million and any budget strictly above that level works at Dealer 1. At the infimum, a low equilibrium still exists.

\begin{table}[htbp]\centering\small
\caption{Equal-budget cash injections in the margin network}\label{tab:targetcash}
\begin{tabular}{@{}lrrr@{}}\toprule
Recipient & Core threshold $Q_j(C)$ & \shortstack{Core fractions at 10.25\\in the least equilibrium} & \shortstack{Early payments\\at 10.25}\\\midrule
Dealer 1 & $10.06730$ & $(1.000,1.000)$ & $255.00$\\
Dealer 2 & $10.64109$ & $(0.000,0.020)$ & $24.36$\\
Correspondent & $\infty$ & $(0.000,0.000)$ & $22.00$\\
Client bank & $\infty$ & $(0.000,0.000)$ & $30.00$\\\bottomrule
\end{tabular}
\par\vspace{3pt}{\footnotesize\emph{Notes.} Amounts in millions of pounds. The finite cut-offs must be strictly exceeded. Gross early payments are evaluated at the least equilibrium.}
\end{table}

\paragraph{When an upstream bank is the better target}
In a three-bank comparison, each dealer owes 80 to the other and 20 outside, with $e_1=28$ and $e_2=29$. Dealers use $\omega_i=20$ and all banks use $\xi=0.25$, $\mu=1.44$ and $\psi=0.75$. Bank 3 owes 5, all to Dealer 1. Its cash of 18 and reserve target of 16 ($\omega_3=16/c$) give a payment of 2 in every baseline equilibrium. Both dealers pay zero in the least equilibrium. The cut-offs for unique full payment in the core are \pounds2.255 million at Bank 3, \pounds2.900 million at Dealer 1 and \pounds2.904 million at Dealer 2. Each pound paid by Bank 3 adds a pound to Dealer 1's cash and lowers its reserve target by $\nu$ pounds. At \pounds2.5 million, only the upstream injection restores the core. Injecting \pounds3 million there also brings Bank 3 to full payment and raises gross early settlement from 2 to 205.

\paragraph{Removing a payment dependency}
In the four-bank network, an outside facility prefunds an amount $0\le a\le L_{ji}$ of a single claim $j\to i$ at par. Bank $i$ receives $a$ in certain cash at the cut-off and the facility becomes the creditor for that amount. Total debtor obligations and each bank's resources at full settlement are preserved. When $a=L_{ji}$ that edge disappears from the active-bank dependency graph. Table~\ref{tab:linkknockout} shows that removing either dealer edge breaks the only strongly connected feedback loop and gives $\rho(M)=0$ and unique full payment. Removing either peripheral edge preserves the three dealer-core equilibria. Prefunding the client's receipt raises its repayment fraction to one in the low and intermediate equilibria. To compare equally sized interventions, prefund 8 on each link in turn. Only the claim on Dealer 2 held by Dealer 1 gives unique full payment at that amount, with $\rho(M)\simeq1.067$. Unique full payment requires prefunding strictly above the respective cut-offs, approximately 7.826 and 8.272 for $2\to1$ and $1\to2$.

\begin{table}[htbp]\centering\small
\caption{Partial prefunding and full link removal in the margin network}\label{tab:linkknockout}
\begin{tabular}{@{}lrrr@{}}\toprule
Payment link & Face amount & \shortstack{Early payments\\with 8 prefunded} & \shortstack{Early payments\\with full prefunding}\\\midrule
Dealer 1 $\to$ Dealer 2 & $100$ & $24.40$ & $255.00$\\
Dealer 2 $\to$ Dealer 1 & $95$  & $255.00$ & $255.00$\\
Correspondent $\to$ Dealer 1 & $10$ & $22.00$ & $22.00$\\
Dealer 2 $\to$ Client & $10$ & $30.00$ & $30.00$\\\bottomrule
\end{tabular}
\par\vspace{3pt}{\footnotesize\emph{Notes.} Amounts in millions of pounds. The columns report gross early payments by the original banks at the least equilibrium, excluding facility advances. Both counterfactuals are computed separately for each link.}
\end{table}

\paragraph{Four interventions on the same network}
In the margin-window network, the cut-offs for unique full early payment are about 1.143 basis points for Dealer 1's funded late charge, 5.515\% of each dealer's outstanding incoming principal sold at par before the funding need, \pounds10.067 million of additional opening cash for Dealer 1 and \pounds7.826 million of prefunding on Dealer 1's claim on Dealer 2.

Late interest strengthens payment discipline and early access to arrears lowers desired reserves. At full early settlement, no late interest is paid and the arrears facility buys no claims. The debtor's pledge and the facility's purchase capacity must still be credible. Direct injections and claim prefunding supply cash at the early cut-off. Prefunding replaces a contingent receipt with certain cash and transfers the claim to the facility. At the respective cut-offs, it uses about 22\% less upfront cash than a direct injection at Dealer 1.

\section{Conclusion}\label{sec:conclusion}
Precautionary reserve demand can sustain a morning freeze even when every obligation is ultimately honoured. The same feedback can amplify delays in a unique equilibrium. At fixed liabilities and opening cash, two reserve parameters determine baseline clearing. The freeze margin uses bank-level inputs to flag an all-zero equilibrium during full settlement, while the network links govern propagation and help determine which partial freezes are possible.

The Basel intraday monitoring framework specifies opening liquidity, timed payments and time-specific obligations \citep{BCBSIntraday2013}. Pillar~3 liquidity disclosures inform margin and rollover exposure scenarios \citep{BCBSLiquidityDisclosures2019}. Matched settlement and funding observations can help locate a network relative to the coexistence threshold. Estimating how missed receipts affect reserve demand requires variation that separates the effect of missed payments from common funding stress (Appendix~\ref{app:identification}).

An upstream injection can end a freeze with less cash when the recipient can pass funding into the core and lower its reserve demand. In the four-bank illustration, the next pound generates the most early settlement at the client bank. Dealer 1 has the lowest cash-injection threshold for ending the freeze.

\begingroup
\small
\setstretch{1.0}
\setlength{\bibsep}{0pt}
\bibliographystyle{plainnat}
\bibliography{references}
\endgroup

\clearpage
\appendix
\titleformat{\section}[block]{\normalfont\normalsize\centering}{Appendix \thesection:}{0.55em}{}
\renewcommand{\theequation}{\thesection.\arabic{equation}}
\setcounter{equation}{0}
\numberwithin{equation}{section}
\section{Proofs and Further Equilibrium Details}\label{app:proofs}
\subsection{Final settlement accounting and the private objective}
\begin{proof}[Proof of Proposition~\ref{prop:terminal}]
Fix $s_{-i}$, write $a=a_i(s_{-i})$, $p=\pb_i s_i$, and note that feasibility implies $0\le p\le\min\{\pb_i,a\}$. Hence $R=a-p\ge0$ and $A^1=\pb_i-p\ge0$. From the terminal rule,
\[
p+Q_i^1=p+\min\{\pb_i-p,a-p+X_i\}=\min\{\pb_i,a+X_i\}.
\]
Similarly, $\mathcal D_i^1=[(\pb_i-p)-(a-p+X_i)]^+=[\pb_i-a-X_i]^+$ and
$R+X_i-Q_i^1=[a+X_i-\pb_i]^+$. At a fixed counterparty profile, the three quantities are independent of $p$ for every $X_i$. Therefore for any measurable, integrable terminal penalty $\mathrm{Pen}_i^{\mathrm{def}}(\mathcal D_i^1)$, its conditional expectation is independent of $s_i$. Subtracting it from $U_i^{\mathrm{net}}$ does not change the bank's best-response set. If $\mathrm{Pen}_i^{\mathrm{def}}$ is nondecreasing and the law of $X_i$ is invariant to counterparty profiles, higher $s_{-i}$ makes $a_i$ weakly larger and $\mathcal D_i^1$ weakly smaller in a common coupling, so the completed payoff retains the monotone indirect-payoff argument for ordered equilibria.
\end{proof}

\begin{proof}[Proof of Proposition~\ref{prop:terminalloss}]
For $s'\ge s''$, the date-0 inflow map $z_i(s_{-i})$ and available resources $a_i(s_{-i})$ are componentwise nondecreasing. The terminal rule gives $\mathcal D_i^1(s,X_i)=[\pb_i-a_i(s_{-i})-X_i]^+$, hence the claimed componentwise ordering for every fixed $X_i\ge0$. In the mechanical comparison, the strategic repayment operator lies weakly below the mechanical operator at each profile; monotone fixed-point iteration therefore orders the corresponding greatest and least equilibria as $s^{\G}\le s^{\EN}$. Applying the first result gives \eqref{eq:terminalorderEN}; summing and dividing by $\pbtot$ preserves the inequality. Finally, since $\pb_i=p_i^0+Q_i^1+\mathcal D_i^1$ by the date-0/terminal accounting identity, summation gives \eqref{eq:terminaldecomposition}.
\end{proof}

\subsection{Feasibility and the private optimisation problem}
\begin{lemma}[Feasible payment sets]\label{lem:feasible}
For any $i\in\A$, $z_i$, $a_i$, and $\bar s_i$ are continuous and componentwise nondecreasing in $s_{-i}$. The correspondence $\mathcal S_i(s_{-i})=[0,\bar s_i(s_{-i})]$ is nonempty, compact valued, continuous, and ascending: if $s_{-i}\le\tilde s_{-i}$, then $\mathcal S_i(s_{-i})\subseteq \mathcal S_i(\tilde s_{-i})$.
\end{lemma}
\begin{proof}
The map $z_i=\sum_{j\ne i}L_{ji}s_j$ has nonnegative coefficients. The positive-part map and $x\mapsto\min\{1,x/\pb_i\}$ on $x\ge0$ preserve continuity and monotonicity. Each feasible set is a closed interval with fixed lower endpoint zero and a continuous upper endpoint in $[0,1]$, giving the stated correspondence properties. The inclusion follows directly from the ordering of the upper endpoints.
\end{proof}

\begin{proof}[Proof of Proposition~\ref{prop:bestresponse}]
By \eqref{eq:ownmarginal}, the objective is strictly concave in $s_i$ over $\mathcal S_i(s_{-i})$, hence admits a unique maximiser. If $\delta_i\ge\chi_i$, the derivative is nonnegative for every feasible reserve $R_i\ge0$ because $\bar G_i(R_i/\kappa_i)\le1$. Therefore the maximum is at $s_i=\bar s_i$. In this case $c_i=H_i=0$.

Now suppose $0<\delta_i<\chi_i$. Since $\bar G_i$ is continuous, strictly decreasing from one to zero, the equation $\bar G_i(R/\kappa_i)=\delta_i/\chi_i$ has the unique solution $R=\kappa_i c_i=H_i>0$. Because $R_i=a_i-\pb_i s_i$, the algebraic unconstrained repayment optimum is $(a_i-H_i)/\pb_i$. Strict concavity implies the constrained optimum is its projection onto $[0,\min\{1,a_i/\pb_i\}]$. Since $H_i>0$, this is $\clip((a_i-H_i)/\pb_i)$. The same formula holds in the case $H_i=0$, giving $B_i=\clip((a_i-H_i)/\pb_i)$ for all parameters.

Inside the clipped expression, $a_i=[e_i+z_i]^+$ can be replaced by $e_i+z_i$. The two agree when $e_i+z_i\ge0$ and both expressions clip to zero when $e_i+z_i<0$, because $H_i\ge0$. Substitute $H_i=h_i+\nu_i(\zb_i-z_i)$ to obtain both forms of \eqref{eq:BR}.
\end{proof}

\begin{proof}[Proof of Proposition~\ref{prop:deviation}]
For $a_i>0$, the upper feasible repayment fraction $\Phi_i=\bar s_i$ is strictly positive. At that upper endpoint, the reserve is $R_i^\Phi=\pos{a_i-\pb_i}$. Differentiability of the objective on the feasible interval and strict concavity imply that its unique maximiser lies strictly below $\Phi_i$ if and only if its left derivative at $\Phi_i$ is strictly negative. From \eqref{eq:ownmarginal} this derivative is $\pb_i[\delta_i-\chi_i \bar G_i(R_i^\Phi/\kappa_i)]$. Thus the strict-deviation criterion is precisely \eqref{eq:deviationtest}. The non-strict reverse inequality makes $\Phi_i$ the unique maximiser. If $a_i=0$, the feasible interval contains only zero, so the wedge vanishes by definition. Since $\bar G_i(x)\le1$, the criterion is impossible if $\delta_i\ge\chi_i$.
\end{proof}

\subsection{Cross-derivatives, optimal retention, and state dependence}
\begin{proof}[Proof of Proposition~\ref{prop:amplification}]
In the interior response region, clipping is inactive. Differentiating \eqref{eq:BR}, using $\partial z_i/\partial s_j=L_{ji}$ and $\partial\Delta_i/\partial s_j=-L_{ji}$, yields
\[
\frac{\partial B_i}{\partial s_j}
=\frac{L_{ji}-\partial H_i/\partial s_j}{\pb_i}
=\frac{(1+\nu_i)L_{ji}}{\pb_i}.
\]
The derivative of money payments with respect to $z_i$, holding the full-payment inflow $\zb_i$ fixed, is $1+\nu_i$. The $m_i$ formula follows from $z_i=\zb_i m_i$ when $\zb_i>0$.

When $R_i>0$, holding $s_i$ fixed gives $\partial R_i/\partial s_j=L_{ji}$ and $\partial\kappa_i/\partial s_j=-\psi_i L_{ji}$. Since $\partial U_i/\partial s_i=\pb_i(\delta_i-V_{i,R})$,
\[
\frac{\partial^2 U_i}{\partial s_i\partial s_j}
=-\pb_i\bigl(V_{i,RR}L_{ji}+V_{i,R\kappa}(-\psi_i L_{ji})\bigr)
=\pb_iL_{ji}(-V_{i,RR}+\psi_i V_{i,R\kappa}).
\]
The formulas for $V_{i,RR}$ and $V_{i,R\kappa}$ in \eqref{eq:Vderiv} give \eqref{eq:crosspartial}. At an optimum strictly inside the feasible interval, $R_i/\kappa_i=c_i$ also gives the response derivative by implicit differentiation. Strict clipping makes the best response locally constant.
\end{proof}

\begin{proof}[Verification of the net-payoff externality, Remark~\ref{rem:netexternality}]
Fix a feasible $s_i$ with $R_i>0$ and write $\mathsf G_i=\kappa_iY_i-R_i$. By the chain rule, for an almost surely nonzero $\mathsf G_i$,
\[
\frac{\partial (\mathsf G_i)^+}{\partial s_j}
=\ind_{\{\mathsf G_i>0\}}\left(-\psi_iL_{ji}Y_i-L_{ji}\right).
\]
Integrability of $Y_i$ permits differentiation inside the conditional expectation under the maintained scale-family distribution. The discipline cost is constant in $s_j$, and the net payoff derivative therefore equals
\[
-\chi_i\EE\left[\ind_{\{\mathsf G_i>0\}}\left(-L_{ji}(1+\psi_iY_i)\right)\right]
=\chi_i L_{ji}\left[\bar G_i(R_i/\kappa_i)+\psi_i\EE\{Y_i\ind_{\{\kappa_iY_i>R_i\}}\}\right].
\]
If $e_i+z_i<0$, the resource derivative is locally zero and the first term does not arise, while a change in the stress scale may still change the expected future shortage. The equality is not asserted at the kink $e_i+z_i=0$.
\end{proof}

\subsection{Existence, monotonicity, and the network multiplier}
\begin{proof}[Proof of Proposition~\ref{prop:lattice}]
By \eqref{eq:matrices}, every entry of $M$ is nonnegative. Clipping an affine map with nonnegative slopes gives a continuous isotone map from the complete lattice $[0,1]^{|\A|}$ into itself. Tarski's fixed-point theorem \citep{Tarski1955} yields a nonempty complete lattice of equilibria. In particular, its greatest and least fixed points exist. Because $B(\one)\le\one$ and $B(0)\ge0$, iterations from $\one$ are nonincreasing and those from zero are nondecreasing. Each bounded monotone sequence converges componentwise. Continuity makes its limit a fixed point. By isotonicity, iterates from $\one$ dominate every fixed point and those from zero are dominated by every fixed point, so the limits are the greatest and least fixed points respectively.

For $s,t\in[0,1]^{|\A|}$, the scalar clip is one-Lipschitz, hence
\begin{equation}\label{eq:lipschitz_app}
|B(s)-B(t)|\le M|s-t|\qquad\text{componentwise}.
\end{equation}
When $\rho(M)<1$, the Neumann series gives $v=(I-M)^{-1}\one\ge\one$. Since $Mv=v-\one$, the constant $\beta=1-1/\|v\|_\infty\in[0,1)$ satisfies $Mv\le\beta v$. For the weighted sup norm $\|x\|_v=\max_i |x_i|/v_i$, inequality \eqref{eq:lipschitz_app} implies $\|B(s)-B(t)\|_v\le\beta\|s-t\|_v$. Banach's fixed-point theorem gives uniqueness and geometric convergence. Finally $M\le(1+\max_i\nu_i)M_0$ entrywise; monotonicity of spectral radius for nonnegative matrices gives the sufficient bound in the proposition.
\end{proof}

\begin{remark}[Bounded linear complementarity formulation]
A profile $s$ is a fixed point if and only if for each active bank exactly one of the following regime conditions holds:
\[
\begin{cases}
s_i=0, &(g+Ms)_i\le0,\\
0<s_i<1, &s_i=(g+Ms)_i,\\
s_i=1, &(g+Ms)_i\ge1
\end{cases}.
\]
These are the box-constrained mixed complementarity conditions for the residual $\mathcal R(s)=(I-M)s-g$. Introducing a multiplier $\lambda_i$ for each upper bound gives a linear complementarity problem in $(s,\lambda)$:
\[
0\le s\perp \mathcal R(s)+\lambda\ge0,\qquad
0\le\lambda\perp\one-s\ge0 .
\]
At zero and interior repayments the multiplier vanishes. At full repayment it equals $-\mathcal R_i(s)\ge0$. For a given $\mathcal Z,\mathcal I,\mathcal F$ partition, \eqref{eq:regimes} computes the interior candidate when the interior matrix is invertible, and the regime inequalities test the candidate.
\end{remark}

\begin{lemma}[Comparative statics of extremal equilibria]\label{lem:monocomp}
If two collections of primitives, holding $L$ and $\pb$ fixed, generate continuous isotone maps $B^{(1)}\le B^{(2)}$ pointwise, their least and greatest equilibria are ordered in the same direction. Consequently, extremal repayments are nonincreasing in $h_i$ and $\nu_i$ and nondecreasing in $e_i$ when other primitives are held fixed.
\end{lemma}
\begin{proof}
Start both maps at zero and iterate; pointwise order and isotonicity imply by induction $B^{(1)k}(0)\le B^{(2)k}(0)$ for every $k$. Limits give $s^{-,(1)}\le s^{-,(2)}$. The same argument starting at $\one$ yields ordering of greatest fixed points. The primitive directions follow from \eqref{eq:B_hnu} and $\Delta_i\ge0$.
\end{proof}

\paragraph{Proof of Proposition~\ref{prop:scc} and Corollary~\ref{cor:acyclic}}
Permuting the rows and columns of the nonnegative matrix $M$ according to the condensation directed acyclic graph of its strongly connected components puts $M$ into Frobenius block triangular form. Its diagonal blocks are precisely $M_{\mathcal C_k\mathcal C_k}$. The eigenvalues of a block triangular matrix are the union of the eigenvalues of its diagonal blocks, so $\rho(M)=\max_k\rho(M_{\mathcal C_k\mathcal C_k})$. If all block radii were strictly below one, then $\rho(M)<1$ and Proposition~\ref{prop:lattice} would give uniqueness. Contraposition proves the necessary condition for multiplicity. An acyclic graph has only singleton SCCs. As $L_{ii}=0$, each diagonal block is the zero matrix, so $\rho(M)=0$ and $M$ is nilpotent after the same permutation. Again Proposition~\ref{prop:lattice} yields uniqueness.

\begin{proof}[Proof of Proposition~\ref{prop:basin}]
The inequalities $1-m<g<0$ imply $b(0)=0$, $b(1)=1$, and $x^*=-g/(m-1)\in(0,1)$ solves $x^*=g+mx^*$. On symmetric profiles, the update reduces to $x_{t+1}=b(x_t)$. For $x_t<x^*$ in the interior regime, $b(x_t)-x_t=g+(m-1)x_t<0$; clipping at zero preserves monotonicity. The decreasing bounded sequence converges to the only fixed point below $x^*$, namely zero. Above $x^*$ the analogous sequence increases to one. Strict clipping at zero and one makes each corner best response locally constant for both banks, establishing local stability to sufficiently small asymmetric perturbations.\qedhere
\end{proof}

\subsection{Coexistence, latent fragility and the benchmark}
\begin{proof}[Proof of Proposition~\ref{prop:coordination}]
Since $z_i(0)=0$, the zero profile is a fixed point exactly when
\[
B_i(0)=\clip\left(\frac{e_i-h_i-\nu_i\zb_i}{\pb_i}\right)=0
\quad\text{for every }i,
\]
which is equivalent to \eqref{eq:zeroeq}. At $s=\one$, the inflow is $\zb_i$ and the confidence shortfall vanishes; thus
\[
B_i(\one)=\clip\left(\frac{e_i+\zb_i-h_i}{\pb_i}\right)=1
\quad\Longleftrightarrow\quad
 e_i+\zb_i\ge\pb_i+h_i.
\]
This proves the full-payment condition. If both inequalities hold the profiles zero and $\one$ are distinct equilibria for $\A\ne\varnothing$. Combining them gives $\pb_i-\zb_i+h_i\le e_i\le h_i+\nu_i\zb_i$, whose width is $(1+\nu_i)\zb_i-\pb_i$. If one condition fails, the corresponding extreme profile is not an equilibrium, so the claimed equivalence is exact for coexistence of \emph{these two profiles}.
\end{proof}

\begin{proof}[Proof of Corollary~\ref{cor:latent}]
The mechanical map is $\Phi_i(s)=\clip((e_i+z_i(s))/\pb_i)$ with Lipschitz exposure matrix $M_0$. Thus $\rho(M_0)<1$ gives a unique mechanical fixed point. The full-payment condition $e_i+\zb_i\ge\pb_i+h_i\ge\pb_i$ implies $\Phi_i(\one)=1$ for every active $i$, so its unique fixed point is $\one$. Since $\Delta_i(\one)=0$, the strategic full-payment condition is independent of $\nu$ and remains satisfied as $\nu$ varies holding $h$ fixed. If the additional zero-repayment inequalities are satisfied, Proposition~\ref{prop:coordination} gives $0$ as a distinct strategic fixed point.
\end{proof}

\begin{proof}[Proof of Corollary~\ref{cor:threshold}]
The all-zero profile is a fixed point if and only if $e_i-h_i\le q\bar\nu_i\zb_i$ for all active $i$. For any institution with positive $\bar\nu_i\zb_i$, this inequality is equivalent to $q\ge(e_i-h_i)/(\bar\nu_i\zb_i)$. If the coefficient vanishes, the inequality is either always satisfied (when $e_i\le h_i$) or never satisfied (when $e_i>h_i$). Taking the maximum over active positive-coefficient constraints, together with $q\ge0$, gives \eqref{eq:threshold}; the maximum over an empty positive-coefficient set is understood as $-\infty$.
\end{proof}

\begin{proof}[Proof of Proposition~\ref{prop:mechanical}]
Since $H_i\ge0$, the monotonicity of scalar clipping implies
\[
B_i(s)=\clip\bigl((e_i+z_i-H_i)/\pb_i\bigr)
\le\clip\bigl((e_i+z_i)/\pb_i\bigr)=\Phi_i(s).
\]
Both operators are isotone. The set of post-fixed points of $B$, $\{x:x\le B(x)\}$, is contained in the analogous set for $\Phi$, so their lattice suprema satisfy $s^+\le s_\Phi^+$. Likewise, $\{x:\Phi(x)\le x\}\subseteq\{x:B(x)\le x\}$, so their lattice infima satisfy $s^-\le s_\Phi^-$. These supremum/infimum characterisations are part of Tarski's theorem. When every $\delta_i\ge\chi_i$, every $c_i=0$, whence $H_i=0$ and the maps coincide. For $H_i>0$, equality $B_i=\Phi_i$ at a specific profile can arise at their common zero boundary ($e_i+z_i\le0$) or full boundary ($e_i+z_i-H_i\ge\pb_i$); thus no global converse follows from equality at one state.
\end{proof}

\begin{proof}[Nonnegativity of the two excess-shortfall components]
For matched greatest or matched least equilibria, Proposition~\ref{prop:mechanical} gives $s^{\EN}\ge s^{\G,0}$ componentwise. Since $RR(s)$ is a nonnegative weighted sum of the actions, the first term in \eqref{eq:ESRdecomp} is nonnegative. Increasing each $\psi_i$ from zero cannot lower the scale pass-through requirement for any given nonnegative $\Delta_i$; it increases $\nu_i=\psi_ic_i$ while leaving $h_i$ unchanged. By Lemma~\ref{lem:monocomp}, matched extremal equilibria weakly decrease, so $s^{\G,0}\ge s^\G$, giving the second nonnegative term.
\end{proof}

\begin{proof}[Proof of Proposition~\ref{prop:pareto}]
Fix $s_{-i}\le\tilde s_{-i}$ and an action $s_i\in \mathcal S_i(s_{-i})$. Lemma~\ref{lem:feasible} implies $s_i\in \mathcal S_i(\tilde s_{-i})$. Improved counterparty repayment weakly raises $a_i$ and hence the reserve $R_i$ for the fixed action, and it weakly lowers $\kappa_i$ because $\Delta_i$ falls. For each fixed realisation $Y_i\ge0$, the shortage $(\kappa_iY_i-R_i)^+$ is nonincreasing when $R_i$ rises and $\kappa_i$ falls. Accordingly $U_i^{\mathrm{net}}(s_i,\tilde s_{-i})\ge U_i^{\mathrm{net}}(s_i,s_{-i})$, with the discipline cost unchanged. Therefore the indirect maximised payoff
\[
\mathcal W_i(s_{-i}):=\max_{s_i\in \mathcal S_i(s_{-i})}U_i^{\mathrm{net}}(s_i,s_{-i})
\]
is nondecreasing in $s_{-i}$. At any strategic equilibrium $s$, the equality $U_i^{\mathrm{net}}(s)=\mathcal W_i(s_{-i})$ holds. Hence if $s'\ge s''$, $U_i^{\mathrm{net}}(s')\ge U_i^{\mathrm{net}}(s'')$ for every active $i$. The proof uses the finite mean of $Y_i$ to ensure net payoffs are defined.
\end{proof}

\subsection{Proof of the arrears-recovery response}\label{app:arrearsproof}
\begin{proof}[Proof of Proposition~\ref{prop:arrears}]
Fix $s_{-i}$ so that $\phi_i\Delta_i$ and $\kappa_i$ are fixed in the bank's own optimisation. The marginal liquidity benefit of increasing $R_i$ is $\chi_i\bar G_i((R_i+\phi_i\Delta_i)/\kappa_i)$. The right derivative is strictly decreasing in $R_i\ge0$. If $\delta_i<\chi_i$ and the critical quantile is $c_i$, the optimal reserve target is
\[
H_i^{\mathrm{rec}}=\big[\kappa_ic_i-\phi_i\Delta_i\big]^+
=\big[h_i+(\nu_i-\phi_i)\Delta_i\big]^+.
\]
When $\delta_i\ge\chi_i$, $c_i=0$ and the target is zero, because $\chi_i\bar G_i(\phi_i\Delta_i/\kappa_i)\le\chi_i\le\delta_i$. Projecting the target-derived payment onto the date-0 budget interval yields \eqref{eq:arrearsBR}; the same positive-part identity used in Proposition~\ref{prop:bestresponse} permits replacing $a_i=[e_i+z_i]^+$ with $e_i+z_i$ inside the clipped expression. If $\phi_i\le\nu_i$, then $H_i^{\mathrm{rec}}\ge0$ is affine in $\Delta_i$, giving the globally clipped-affine form and the derivative \eqref{eq:arrearsSlope}. For arbitrary $\phi_i\in[0,1]$, wherever $H_i^{\mathrm{rec}}>0$, increasing $z_i$ raises $e_i+z_i-H_i^{\mathrm{rec}}$ with derivative $1+\nu_i-\phi_i\ge0$. On intervals where $H_i^{\mathrm{rec}}$ is identically zero, the slope equals one. The pieces meet continuously at the reserve target's kink. Consequently the target-adjusted repayment expression is continuous and nondecreasing in every counterparty payment, and clipping preserves both properties.
\end{proof}

\section{Eisenberg--Noe Clearing and the Role of Contraction}\label{app:en}
For nonnegative external resources, the Eisenberg--Noe payment operator is
\[
T(p)=\min\{\pb,e+\Pi^\top p\},\qquad 0\le p\le\pb,
\]
with the minimum taken componentwise. Clearing requires $p=T(p)$. Theorem~1 of \citet{EisenbergNoe2001} establishes extremal clearing vectors using monotonicity and a lattice fixed-point argument; their Theorem~2 establishes uniqueness under regularity. Neither result requires the full clearing operator to be a strict contraction. On active coordinates, substituting $p_i=\pb_i s_i$ gives the mechanical operator $\Phi$ in Definition~\ref{def:mechanical}.

A two-bank example separates uniqueness from contraction. Let $L_{12}=L_{21}=1$ and $e_1=e_2=\varepsilon\in(0,1)$. Then
\[
\Phi(s)=\bigl(\min\{1,\varepsilon+s_2\},
                  \min\{1,\varepsilon+s_1\}\bigr).
\]
Its unique fixed point is $(1,1)$: both coordinates cannot be below one because their fixed-point equations would imply $0=2\varepsilon$, and either coordinate equal to one forces the other to equal one. Nevertheless, $\rho(M_0)=1$. Along $s=(t,t)$ for $0\le t<1-\varepsilon$, the map preserves distances, so it is not a strict contraction in any norm.

In the strategic model, existence and extremal-equilibrium iteration likewise hold without contraction by Proposition~\ref{prop:lattice}. Contraction supplies an additional, checkable \emph{sufficient} condition for global uniqueness and geometric convergence from every initial profile. Specifically,
\[
|B(s)-B(t)|\le M|s-t|,\qquad M=(I+D_\nu)M_0.
\]
When $\rho(M)<1$, this bound yields a strict contraction in a weighted sup norm, as proved in Appendix~\ref{app:proofs}. The matrix $M$ records the cash-flow effect of payment links and the additional reserve response; it bounds feedback across all clipping regimes. Its spectral radius need not be below one for a particular equilibrium to be unique, since clipping can attenuate the realised response. The criterion also makes the comparison transparent: obligations leading to pure creditors can make $\rho(M_0)<1$, while confidence amplification can raise $\rho(M)$ above one. Thus contraction identifies a region of strategic robustness; it is not an assumption needed to define or establish the existence of either clearing mechanism.

\section{Distributional Models: Full GPD and Peaks over Threshold}\label{app:distribution}
\subsection{Full-distribution generalised Pareto specification}
The general theory in the main text uses the scale-family assumptions on the nonnegative random variable $Y_i$. The illustrations specify its survivor as $\bar G_i(y)=S_{\xi_i}(y)$, where
\begin{equation}\label{eq:GPDsurv}
S_{\xi_i}(y)=\begin{cases}
(1+\xi_i y)^{-1/\xi_i},&\xi_i>0,\\
e^{-y},&\xi_i=0
\end{cases}\qquad y\ge0,
\end{equation}
Here $0\le\xi_i<1$ ensures $\EE[Y_i]=(1-\xi_i)^{-1}<\infty$. Conditional on $s_{-i}$, the future need $\Lambda_i=\kappa_iY_i$ has scale $\kappa_i(s_{-i})$ and shape $\xi_i$. The density $f_i$ for this specification is
\begin{equation}\label{eq:GPDdensity}
f_i(y)=\begin{cases}
(1+\xi_i y)^{-1/\xi_i-1},&\xi_i>0,\\
e^{-y},&\xi_i=0
\end{cases}.
\end{equation}
For $\mu_i:=\chi_i/\delta_i>1$, the optimal dimensionless retention quantile is
\begin{equation}\label{eq:GPDquantile}
c_i=\begin{cases}
\displaystyle\frac{\mu_i^{\xi_i}-1}{\xi_i},&\xi_i>0,\\[6pt]
\log\mu_i,&\xi_i=0
\end{cases},
\end{equation}
with $c_i=0$ when $\mu_i\le1$. Thus $c_i$ is weakly increasing in $\chi_i/\delta_i$, with a strict increase for ratios above one. For $\mu_i>1$, it is also increasing in the GPD shape $\xi_i$; an exact derivative at $\xi_i>0$ is
\[
\frac{\partial c_i}{\partial \xi_i}
=\frac{1+\mu_i^{\xi_i}(\xi_i\log\mu_i-1)}{\xi_i^2}>0,
\qquad \lim_{\xi_i\downarrow0}\frac{\partial c_i}{\partial\xi_i}=
\frac12(\log\mu_i)^2.
\]
This comparative static concerns the retention quantile within the scale family; counterparty non-payment probabilities are separate objects.

The full-distribution GPD admits explicit liquidity values. For $0<\xi_i<1$ and $R\ge0$,
\begin{align}
V_i(R;\kappa)&=\frac{\chi_i\kappa}{1-\xi_i}
\left[1-\left(1+\xi_i\frac{R}{\kappa}\right)^{1-1/\xi_i}\right],\label{eq:GPDgross}\\
V_i^{\mathrm{net}}(R;\kappa)&=-\frac{\chi_i\kappa}{1-\xi_i}
\left(1+\xi_i\frac{R}{\kappa}\right)^{1-1/\xi_i}.
\label{eq:GPDnet}
\end{align}
The limit $\xi_i=0$ gives $V_i=\chi_i\kappa(1-e^{-R/\kappa})$ and $V_i^{\mathrm{net}}=-\chi_i\kappa e^{-R/\kappa}$. Because $\EE[\Lambda_i\mid s_{-i}]=\kappa_i(s_{-i})/(1-\xi_i)$ depends on counterparty repayments, only the net expression is appropriate for comparing different equilibrium states.

For $\xi_i>0$, the conditional law of $\Lambda_i$ is also a Lomax distribution with shape $1/\xi_i$ and scale $\kappa_i(s_{-i})/\xi_i$. The GPD parameterisation makes the extreme-value shape explicit and includes the exponential boundary.

\subsection{Peaks-over-threshold specification}
Classical POT analysis models the \emph{conditional positive excess} of a random total requirement $\widetilde\Lambda_i$ above a sufficiently high threshold $u_i$. For a given counterparty profile,
\begin{equation}\label{eq:POTpositive}
\widetilde\Lambda_i-u_i\ \big|\ \widetilde\Lambda_i>u_i,\,s_{-i}\ \sim\ \mathrm{GPD}(\xi_i,\kappa_i(s_{-i})).
\end{equation}
Exceedance limits motivate this approximation \citep{BalkemaDeHaan1974,Pickands1975}. Estimating $\PP(\widetilde\Lambda_i>R)$ below $u_i$ requires information on the distribution below the threshold.

An EVT-compatible \emph{alternative economic object} is the extra funding need \emph{above} a stress threshold,
\begin{equation}\label{eq:POTlambda}
\Lambda_i=(\widetilde\Lambda_i-u_i)^+.
\end{equation}
If the conditional exceedance probability $q_i^{\mathrm{POT}}=\PP(\widetilde\Lambda_i>u_i\mid s_{-i})\in(0,1]$ is fixed across repayment profiles, and the conditional positive excess is GPD with scale $\kappa_i(s_{-i})$, then $\Lambda_i$ has a point mass $1-q_i^{\mathrm{POT}}$ at zero and, for $R\ge0$,
\begin{equation}\label{eq:POTsurv}
\PP(\Lambda_i>R\mid s_{-i})
=q_i^{\mathrm{POT}}\left(1+\xi_i\frac{R}{\kappa_i(s_{-i})}\right)^{-1/\xi_i},
\end{equation}
with the exponential limit for $\xi_i=0$. Conditional on the payment profile, its mean is $q_i^{\mathrm{POT}}\kappa_i/(1-\xi_i)$ for $0\le\xi_i<1$. The atom at zero when $q_i^{\mathrm{POT}}<1$ changes the right derivative of liquidity value to $V_{i,R}(0+)=\chi_iq_i^{\mathrm{POT}}$.

For $R>0$ the marginal liquidity-insurance value is now $\chi_i q_i^{\mathrm{POT}}S_{\xi_i}(R/\kappa_i)$. If $\delta_i<\chi_iq_i^{\mathrm{POT}}$, the desired reserve is $H_i=\kappa_ic_i^{\mathrm{POT}}$ with
\begin{equation}\label{eq:POTcritical}
c_i^{\mathrm{POT}}=\begin{cases}
\displaystyle\frac{(\chi_iq_i^{\mathrm{POT}}/\delta_i)^{\xi_i}-1}{\xi_i},&\xi_i>0,\\[6pt]
\log(\chi_iq_i^{\mathrm{POT}}/\delta_i),&\xi_i=0
\end{cases}.
\end{equation}
Set $c_i^{\mathrm{POT}}=0$ when $\delta_i\ge\chi_iq_i^{\mathrm{POT}}$. Since $q_i^{\mathrm{POT}}$ is constant in the bank's counterparty environment in this alternative, $c_i^{\mathrm{POT}}$ is also constant, and \emph{the same clipped-affine best-response structure survives} with $c_i$ replaced by $c_i^{\mathrm{POT}}$. The projection argument in Appendix~\ref{app:proofs} then gives the best response. The baseline financial plan covers funding needs up to the threshold.

A payment-dependent $q_i^{\mathrm{POT}}$ can make the critical ratio vary with the payment profile even at a fixed scale. The clearing rule is then generally non-affine.

\subsection{Estimation and validation implications}
Fit the full-distribution model to the complete sample of funding needs, conditional on the funding state. For POT, estimate the excess distribution from positive exceedances and the frequency $q_i^{\mathrm{POT}}$ from the full observation window. Check the distributional fit at the settlement horizon and account for dependence across funding events.

\section{Identification and Measurement}\label{app:identification}
\subsection{Observed or constructed financial quantities}
Opening usable cash and dated payment records can inform $e_i$, scheduled exposures and missed receipts. The Basel intraday monitoring framework uses opening liquidity, time-stamped flows and time-specific obligations \citep{BCBSIntraday2013}. Construct unrestricted settlement cash separately from pledged cash and borrowing capacity. Where bilateral records are unavailable, observed bank totals can constrain the reconstruction of $L$.
\begin{table}[htbp]\centering\footnotesize
\caption{Financial data objects: observability and construction}\label{tab:data}
\begin{tabularx}{\linewidth}{@{}l>{\raggedright\arraybackslash}X>{\raggedright\arraybackslash}X@{}}\toprule
Object & Economic interpretation & Potential data status \\\midrule
$L_{ij}$ & Directed bilateral contractual obligation & Bilateral records if accessible; otherwise \emph{reconstructed} network conditional on total constraints.\\
$\pb_i$ & Total modelled outgoing interbank obligations & Regulatory aggregate, if consistently defined; otherwise row sum of $L$.\\
$\zb_i$ & Total incoming contractual interbank claims & Interbank asset aggregate where available; otherwise column sum of $L$.\\
$e_i$ & Date-0 net unrestricted external liquidity & Available liquid assets net of immediate outside obligations and excluding the late-interest pledge.\\
Deposits & Uninsured share and total deposit funding & Regulatory balance-sheet and deposit disclosures can inform the risk of future withdrawals.\\
$X_i$ & Date-1 net continuation cash for principal & Cash available at date 1 after senior external claims, excluding the pledge and same-round receipts.\\
$C_i$ & Cash restricted to late-interest payments & Pledge records or restricted-cash disclosures matched to the relevant contracts.\\
\bottomrule
\end{tabularx}
\end{table}

\subsection{Estimated, calibrated, and sensitivity primitives}
Table~\ref{tab:parameters} summarises calibration and empirical requirements.
\begin{table}[htbp]\centering\footnotesize
\caption{Behavioural and funding primitives: proposed treatment}\label{tab:parameters}
\begin{tabularx}{\linewidth}{@{}l>{\raggedright\arraybackslash}X>{\raggedright\arraybackslash}X@{}}\toprule
Primitive & Economic meaning & Suggested evidentiary status \\\midrule
$\omega_i$ & Baseline scale of future additional funding need & Estimate from interim funding outflows at full incoming settlement under the chosen shock normalisation.\\
$\xi_i$ & GPD shape in illustrative full-distribution model & Estimate from total funding needs for the full-distribution specification.\\
$\psi_i$ & Scale response to missing counterparty settlement & Use linked payment and funding records to estimate $\eta_i$. For the GPD, $\psi_i=(1-\xi_i)\eta_i$.\\
$\chi_i$ & Marginal future funding shortfall cost & Match refinancing spreads or liquidation discounts to the funding channel.\\
$r_i^{\mathrm{late}},\tau_i^d$ & Annualised late rate and delay window in years & Contractual rate, day-count convention and time between settlement cut-offs.\\
$d_i^{\mathrm{priv}}$ & Private cost per unit deferred & Estimate or proxy from franchise and funding consequences of deferral.\\
$F_i$ & Franchise value at risk under \eqref{eq:franchise} & Valuation proxy for franchise value exposed to payment delay.\\
$\mathrm{Pen}_i^{\mathrm{def}}(\cdot)$ & Additional penalty for terminal defaulted principal & Calibrated legal/restructuring or franchise costs, distinct from prompt-payment discipline.\\
$\phi_i$ & Fraction of outstanding incoming principal sold at par before interim liquidity stress & Timing-sensitivity scenario; requires dated payment-event records to estimate.\\
$h_i,\nu_i$ & Reduced-form retention intercept and amplification & Sufficient statistics for early repayment. Identifying the structural primitives requires separate evidence on funding needs and costs.\\\bottomrule
\end{tabularx}
\end{table}

\subsection{Behavioural sufficient statistics and structural primitives}
With a fixed liability matrix $L$ and external positions $e$, the clipped-affine best response is determined by
\begin{equation}\label{eq:reducedform}
(h_i,\nu_i)=(\omega_ic_i,\psi_ic_i),
\qquad c_i\text{ as in }\eqref{eq:ci}.
\end{equation}
At fixed $L$ and $e$, calibrations that produce the same baseline reserve $h_i$ and reserve sensitivity $\nu_i$ imply the same early repayment rule and equilibrium set. A stress test can therefore compare alternative funding models through the reserve parameters they imply. With observations of usable cash, missed receipts and repayments, interior choices imply $a_i-p_i=h_i+\nu_i\Delta_i$. This can discipline the reserve schedule directly, subject to simultaneous determination of payments and censoring at zero and full repayment. Separate evidence is needed to identify the funding distribution and cost primitives. Multiplying both costs $\chi_i$ and $\delta_i$ by the same positive factor leaves $c_i$ unchanged and scales net payoffs in \eqref{eq:netU}.

Choose proxies for the funding channel and horizon of interest. For collateral stress, dated margin calls measure interim needs and disclosed derivatives and collateral outflows indicate exposure. For funding rollovers, maturities and renewal records anchor the refinancing gap, while short-dated wholesale funding and funding concentration indicate vulnerability. Pillar 3 liquidity disclosures provide relevant exposure categories \citep{BCBSLiquidityDisclosures2019}. Their reported 30-day stress measures can inform scenario design. Estimating a one-day $\eta_i$ requires matched observations of missed receipts and distinct subsequent funding needs, with variation that separates the response to a missed receipt from common stress. Exclude the missed receipt itself from the subsequent funding need. Where data support a range, compare that range with the coexistence threshold.

Refinancing spreads or liquidation discounts inform $\chi_i$. Contract terms and settlement times determine $k_i$. Estimating $d_i^{\mathrm{priv}}$ requires separate evidence on the franchise or market-access consequences of deferral. Carry-based calibration uses returns on retained cash and effective holding intervals while counting each funding cost once. Report uncertainty in measured or reconstructed bilateral liabilities and net unrestricted liquidity.

\subsection{Interbank exposure versus interbank obligations}

When $\zb_i>0$, incoming claims govern reserve sensitivity to the unpaid share through $\partial H_i/\partial(1-m_i)=\nu_i\zb_i$. Outgoing obligations $\pb_i$ scale repayment fractions and enter the franchise-cost specification in \eqref{eq:franchise}. A network change must trace both directions and its effects on $(e,F,\delta)$ where relevant. The conditional scale equation specifies responses to conjectured payments on fixed links; counterparty default probabilities and network formation require separate modelling.

\section{Nonlinear Confidence as a Robustness Extension}\label{app:nonlinear}
A nondecreasing nonlinear funding scale allows the strength of payment amplification to vary with missed receipts. Suppose the conditional scale is
\begin{equation}\label{eq:nonlinearscale}
\kappa_i(s_{-i})=\omega_i+\psi_i \mathfrak h_i(\Delta_i(s_{-i})),
\end{equation}
where $\mathfrak h_i:[0,\zb_i]\to\RR_+$ is continuous, nondecreasing, $\mathfrak h_i(0)=0$, and differentiable with $0\le \mathfrak h_i'(x)\le K_i^{\mathfrak h}<\infty$. This retains the same future-liquidity distributional form and discipline cost; the only change is the reduced-form response of future stress to the current payment shortfall. The individual retention target is $H_i(s_{-i})=c_i\kappa_i(s_{-i})$, and
\begin{equation}\label{eq:nonlinearBR}
B_i(s_{-i})=\clip\left(\frac{e_i+z_i-\omega_ic_i-\psi_ic_i\mathfrak h_i(\zb_i-z_i)}{\pb_i}\right).
\end{equation}
The joint repayment map $B$ remains continuous and isotone, so its fixed points still form a nonempty complete lattice. In the interior regime,
\begin{equation}\label{eq:nonlinearslope}
\frac{\partial B_i}{\partial s_j}
=\frac{L_{ji}}{\pb_i}\left[1+\psi_ic_i \mathfrak h_i'(\Delta_i)\right].
\end{equation}
Thus counterparties' repayments can affect not only the size of retention but also its \emph{marginal response}. By the bounded derivative assumption, a global Lipschitz comparison matrix has entries
\begin{equation}\label{eq:M_K}
\widetilde M_{ij}=\frac{L_{ji}}{\pb_i}(1+\psi_ic_iK_i^{\mathfrak h}),\qquad i,j\in\A,
\end{equation}
so $\rho(\widetilde M)<1$ is a sufficient uniqueness condition by the same weighted-contraction argument.

One illustrative convex confidence function, for $\zb_i>0$, is
\begin{equation}\label{eq:convexf}
\mathfrak h_i(\Delta_i)=\Delta_i+\alpha_i\frac{\Delta_i^2}{\zb_i},\qquad\alpha_i\ge0.
\end{equation}
Then $\mathfrak h_i'(\Delta_i)=1+2\alpha_i(1-m_i)$, and the interior payment amplification becomes
\begin{equation}\label{eq:stateAmplification}
\frac{\partial(\pb_iB_i)}{\partial z_i}
=1+\nu_i\bigl[1+2\alpha_i(1-m_i)\bigr].
\end{equation}
When $\alpha_i\nu_i>0$, the reserve response to an additional missed pound is larger at higher shortfalls. The accounting derivative $\partial R_i/\partial s_j=L_{ji}$ still holds at fixed $s_i$ when $e_i+z_i>0$, and is zero below the cash kink. Set $\mathfrak h_i\equiv0$ when $\zb_i=0$.

\paragraph{State-dependent shortage prices}
If $\chi_i(\Delta)>0$ is continuous and nondecreasing, with $\delta_i>0$ fixed, define $c_i(\Delta)=0$ when $\chi_i(\Delta)\le\delta_i$ and $c_i(\Delta)=\bar G_i^{-1}(\delta_i/\chi_i(\Delta))$ otherwise. Then $H_i(\Delta)=(\omega_i+\psi_i\Delta)c_i(\Delta)$ is nondecreasing, and $B_i=\clip((e_i+z_i-H_i(\zb_i-z_i))/\pb_i)$ remains continuous and isotone, preserving the equilibrium lattice. If $H_i$ is Lipschitz with constant $K_i$, the matrix with entries $(1+K_i)L_{ji}/\pb_i$ yields a sufficient spectral uniqueness test. The zero- and full-payment conditions of Proposition~\ref{prop:coordination} become $e_i\le (\omega_i+\psi_i\zb_i)c_i(\zb_i)$ and $e_i+\zb_i\ge\pb_i+\omega_i c_i(0)$, respectively. Proposition~\ref{prop:pareto}'s private-payoff ordering persists: higher counterparty payments weakly expand feasible reserves and weakly lower both $\kappa_i$ and $\chi_i$. Joint calibration with $\psi_i$ is needed to avoid double-counting stress.

\section{Numerical and Reproducibility Notes}\label{app:numerics}
\begingroup\small
\subsection{Margin-window thresholds and numerical checks}\label{app:margincalcs}
For the dealers in Section~\ref{subsec:latentex}, $(\gamma_1,\gamma_2)\simeq(-0.024674,-0.073518)$ and $(\beta_1,\beta_2)\simeq(1.062625,1.169396)$ in \eqref{eq:latentBR}. Since $\gamma_i<0$ and $\gamma_i+\beta_i>1$, a boundary fraction requires the same corner at the other dealer. The interior equations give $s_1^*=(\gamma_1+\beta_1\gamma_2)/(1-\beta_1\beta_2)$ and $s_2^*=\gamma_2+\beta_2s_1^*$, both strictly interior. The map $T=b_1\circ b_2$ is increasing and has only the fixed points $0$, $s_1^*$ and $1$. It is below the diagonal just to the right of zero and above the diagonal just to the left of one, giving the stated subsequence limits. Table~\ref{tab:marginoutcomes} uses gross early payment $115s_1+110s_2+10+\min\{20,12+10s_2\}$. The sum of the two dealers' net payoffs in \eqref{eq:netU} is $-69.60$, $-43.42$ and $-7.99$ thousand pounds at the low, intermediate and high equilibria, respectively.

For policy comparisons with $\gamma_2<0$ and $\gamma_i+\beta_i>1$, put $\ell=-\gamma_2/\beta_2$. If $\gamma_1\le\ell$, the core has a low fixed point $(\max(0,\gamma_1),0)$. For $\gamma_1>\ell$, $T(x)>x$ on $[0,\ell]$. On $[\ell,(1-\gamma_2)/\beta_2]$ the unclipped composition is affine and above the diagonal at both endpoints. Clipping it at one preserves the strict inequality for $x<1$. At larger $x$ the composition equals one. Hence full payment is unique precisely when $\gamma_1>\ell$ in this comparison.

As $k_1$ rises with $d_1^{\mathrm{priv}}$ fixed, the zero-core boundary is $\gamma_1=0$ and the uniqueness boundary is $\gamma_1=\ell$. For either target $t\in\{0,\ell\}$, the critical ratio is $c_1=(32-115t)/(20+0.75\cdot95)$ and the charge is $\chi(1+0.25c_1)^{-4}-d_1^{\mathrm{priv}}$. This gives $k_1$ of 1.03846 and 1.14293 basis points, respectively. With common recovery $\phi$ and the original charges, the same boundaries use $\gamma_i(\phi)=\gamma_i+\phi n_i/\pb_i$ and $\beta_i(\phi)=\beta_i-\phi n_i/\pb_i$, where $(n_1,n_2)=(95,100)$ are each dealer's claims on the other dealer. Solving $\gamma_1(\phi)=0$ and $\gamma_1(\phi)=-\gamma_2(\phi)/\beta_2(\phi)$ gives $0.029868$ and $0.055153$. The feedback radius is $\sqrt{\beta_1(\phi)\beta_2(\phi)}$. Beyond the uniqueness threshold, the maps only increase as recovery rises and full payment remains an equilibrium.

\subsection{Topology and targeting calculations}
For Table~\ref{tab:topology}, both networks have $g\simeq-0.015207$ and a common strategic row sum $m\simeq1.028851$. In the diversified network, if $s_i>s_j$, bank $i$'s unclipped response is smaller than bank $j$'s, which rules out a nonuniform fixed point. In the concentrated network put $A=0.895(1+\nu)$ and $D=0.005(1+\nu)$. For within-pair gaps $d_1=s_1-s_2$, $d_2=s_3-s_4$, the corresponding unclipped response gaps are $-Ad_1-Dd_2$ and $-Ad_2+Dd_1$. By monotonicity of clipping, $d_1>0$ forces $d_2<0$, which in turn forces the second response gap to be positive, a contradiction. Reversing signs or taking $d_1=0$ gives $d_1=d_2=0$ at any equilibrium.

The pair-uniform map sends $(u,v)$ to $(\clip(g+Au+Dv),\clip(g+Av+Du))$. Since $g+D<0$ and $g+A>1$, each pair's corner is a best response to any fraction of the other pair. Against the other pair fixed at $v\in\{0,1\}$ the interior threshold is $-(g+Dv)/(A-1)$. With both pairs interior, $A-D>1$ implies they are equal, giving the common threshold $-g/(m-1)$. These cases give the nine equilibria. From the table's start, the diversified iterates are all below $x^*$ at step 2. Monotonicity bounds them above by a uniform trajectory converging to zero. At step 23 the concentrated iterates satisfy $u\ge0.66$ and $v\le0.40$. The first pair then grows towards one by comparison with $u\mapsto\clip(g+Au)$, and the second falls to zero by comparison with $v\mapsto\clip(g+Av+D)$.

In the margin network, write $t_i=\gamma_i+q_i/\pb_i$ for the dealer intercepts after cash injections. Full payment is unique exactly when
\begin{equation}\label{eq:targetcondition}
\max\{t_1+t_2/\beta_2,\ t_2+t_1/\beta_1\}>0 .
\end{equation}
Both slopes exceed one. For mixed intercept signs, the larger expression is the condition from Appendix~\ref{app:margincalcs} for the dealer with the nonnegative intercept. When both intercepts are nonnegative with at least one positive, $b_1\circ b_2$ has positive affine intercept and slope $\beta_1\beta_2>1$ until upper clipping, so it is above the diagonal for $x<1$. When both intercepts are nonpositive, zero remains a core equilibrium. Thus the two single-dealer cut-offs are $\pb_1(-\gamma_2/\beta_2-\gamma_1)$ and $\pb_2(-\gamma_1/\beta_1-\gamma_2)$. The first expression in \eqref{eq:targetcondition} rises more per pound at Dealer 1 than at Dealer 2, since $\beta_2\pb_2>\pb_1$. The second rises more per pound at Dealer 2, since $\beta_1\pb_1>\pb_2$. Consequently, no split can reduce the budget infimum below the smaller single-dealer cut-off. At $q_2=10.25$ the least core equilibrium is $(0,t_2)$ with $t_2\simeq0.019664$, making gross payment $22+120t_2$.

In the three-bank network, $s_3=\clip((e_3-16)/5)$ and the dealer slopes are both $m=0.8(1+\nu)\simeq1.029068$. With Bank 3 paying 2, the core intercepts are $100\gamma_1=28-h-85\nu+2(1+\nu)$ and $100\gamma_2=29-h-80\nu$. The single-dealer cut-offs from \eqref{eq:targetcondition} are $100(-\gamma_2/m-\gamma_1)\simeq2.90031$ and $100(-\gamma_1/m-\gamma_2)\simeq2.90429$. For $q\in[0,3]$ at Bank 3, Dealer 1's intercept rises by $(1+\nu)q/100$. Thus the upstream cut-off is the first direct cut-off divided by $1+\nu$, about $2.25471$.

Prefunding $a$ on $j\to i$ shifts $e_i$ up by $a$ and reduces $L_{ji}$ and $\zb_i$ by $a$, with $\pb_j$ held fixed through assignment. Thus $g_i$ rises and $M_{ij}$ falls by the same amount $(1+\nu_i)a/\pb_i$. On a dealer-to-dealer link the paying dealer's response is unchanged. The uniqueness criterion of Appendix~\ref{app:margincalcs} therefore gives a cut-off of $Q_i(C)/(1+\nu_i)$ for prefunding the receiving dealer's claim. Prefunding the claim on the correspondent leaves receipts unchanged at every fixed point because that bank pays in full. Prefunding the client bank's receipt only changes its own payment.

\subsection{Capacity-constrained calculations and implementation}
For Section~\ref{subsec:nonmultiple}, $s_1=s_6=1$ and $s_5=44/45$ in both clearing regimes. The mechanical fractions for banks 2--4 are $(1,67/80,49/55)$. In the strategic regime solve
\begin{align*}
180s_2&=94.5+59.92s_3+36.38s_4,\\
80s_3&=8.6+42s_2,\\
55s_4&=19.325+27.675s_2 .
\end{align*}
The solution $(0.873213,0.565937,0.790749)$ satisfies these repayment regimes. The interior set for the multiplier is $\mathcal I=\{2,3,4,5\}$. Bank 6 is at its full-payment threshold and receives network payments only from Dealer 1, whose full payment persists under the reported marginal injections. Applying $\Phi$ to this strategic vector gives the feasible-payment column in Table~\ref{tab:nonmultiple}. The row vector $\pb_{\mathcal I}^\top(I-M_{\mathcal I\mathcal I})^{-1}D_{\pb,\mathcal I}^{-1}$ gives the marginal gross-payment effects reported in Section~\ref{subsec:nonmultiple}.

Construct $(\pb,\zb,c,h,\nu,g,M)$ with $L_{ij}$ owed by $i$ to $j$. Iterate $B$ from $0$ and $\one$ to obtain extremal equilibria. Check feasibility, iterate differences and the residual $\|s-B(s)\|_\infty$. Alternatively solve \eqref{eq:regimes} and check the clipping inequalities. Match extremal selections across counterfactuals.
\endgroup

\end{document}